\documentclass[
    reprint,            % Mimics the final two-column journal layout
    aps,                % American Physical Society styling
    prl,                % Physical Review Letters specific formatting
    superscriptaddress, % Use superscripts for authors with different affiliations
    amsmath,            % Essential math environments
    amssymb,            % Standard symbols
    floatfix            % Helps fix float placement issues
]{revtex4-2}
\usepackage{graphicx}                                   % Figure handling
\usepackage[table]{xcolor}                              % Color support (supersedes color)
\usepackage{dcolumn}                                    % Align table columns on decimal point
\usepackage{bm}                                         % Bold math symbols
\usepackage[colorlinks=true, allcolors=blue]{hyperref}  % Clickable links/citations
\usepackage{siunitx}                                    % Professional unit typesetting
\usepackage{mathtools}                                  % Extensions to amsmath
\usepackage{booktabs}                                   % Professional tables
\usepackage[normalem]{ulem}
\usepackage{makecell}
\newcommand{\graynote}[1]{\\[-1mm]{\color{gray}\scriptsize #1}}
\DeclareSIUnit\Gauss{G}
\DeclareSIUnit\rad{rad}
\DeclareSIUnit\mrad{mrad}
\DeclareSIUnit\Eris{\text{$\mathrm{E}_{\ensuremath{i},\mathrm{S}}^{\mathrm{R}}$}}
\DeclareSIUnit\Eril{\text{$\mathrm{E}_{\ensuremath{i},\mathrm{L}}^{\mathrm{R}}$}}
\DeclareSIUnit\Erxs{\text{$\mathrm{E}_{\ensuremath{x},\mathrm{S}}^{\mathrm{R}}$}}
\DeclareSIUnit\Erys{\text{$\mathrm{E}_{\ensuremath{y},\mathrm{S}}^{\mathrm{R}}$}}
\DeclareSIUnit\Erxl{\text{$\mathrm{E}_{\ensuremath{x},\mathrm{L}}^{\mathrm{R}}$}}
\DeclareSIUnit\Eryl{\text{$\mathrm{E}_{\ensuremath{y},\mathrm{L}}^{\mathrm{R}}$}}
\DeclareSIUnit\Erz{\text{$\mathrm{E}_{\ensuremath{z}}^{\mathrm{R}}$}}
\DeclareSIUnit\as{\text{$a_\mathrm{s}$}}
\DeclareSIUnit\ab{\text{$a_\mathrm{B}$}}

\newcommand{\beq}{\begin{equation}}
\newcommand{\eeq}{\end{equation}}
\newcommand{\beqn}{\begin{eqnarray}}
\newcommand{\eeqn}{\end{eqnarray}}
\newcommand{\llangle}{\langle\!\langle}
\newcommand{\rrangle}{\rangle\!\rangle}

\newcommand{\ra}{\rightarrow}
\newcommand{\tr}{\mathrm{tr}}
\newcommand{\vect}[1]{{\bm{#1}}}

\newcommand{\tabref}[1]{Tab.~\ref{#1}}

\begin{document}
\setcounter{secnumdepth}{3}
% =============================================================
% Title and authors
% =============================================================

%%% Add affiliations to the custom \maketitle command
\newcommand{\MPQ}{Max-Planck-Institut f\"{u}r Quantenoptik, 85748 Garching, Germany}
\newcommand{\MCQST}{Munich Center for Quantum Science and Technology, 80799 Munich, Germany}
\newcommand{\LMU}{Fakult\"{a}t f\"{u}r Physik, Ludwig-Maximilians-Universit\"{a}t, 80799 Munich, Germany}
\newcommand{\UCSB}{Department of Physics, University of California, Santa Barbara, California 93106, USA}
\newcommand{\KITP}{Kavli Institute for Theoretical Physics, University of California, Santa Barbara, CA 93106, USA}
\title{Observation of Strong-to-Weak Spontaneous Symmetry Breaking \\ in a Dephased Fermi Gas}

\author{Si Wang}%
\altaffiliation{These authors contributed equally}
\affiliation{\MPQ}%
\affiliation{\MCQST}%
\author{Thomas G. Kiely}
\altaffiliation{These authors contributed equally}
\affiliation{\KITP}
\author{Dorothee Tell}%
\affiliation{\MPQ}%
\affiliation{\MCQST}%
\author{Johannes Obermeyer}%
\affiliation{\MPQ}%
\affiliation{\MCQST}%
\author{Marnix Barendregt}%
\affiliation{\MPQ}%
\affiliation{\MCQST}%
\author{Petar Bojovi\'c}%
\affiliation{\MPQ}%
\affiliation{\MCQST}%
\author{Philipp M. Preiss}%
\affiliation{\MPQ}%
\affiliation{\MCQST}%
\author{Abhijat Sarma}
\affiliation{\UCSB}
\author{Titus Franz}%
\affiliation{\MPQ}%
\affiliation{\MCQST}%
\author{Matthew P.\,A. Fisher}
\affiliation{\UCSB}
\author{Cenke Xu}
\affiliation{\UCSB}
\author{Immanuel Bloch}%
\affiliation{\MPQ}%
\affiliation{\MCQST}%
\affiliation{\LMU}%

\date{\today}

% =============================================================
% Abstract
% =============================================================

\begin{abstract}

Symmetry-based classification of quantum phases of matter is one of the most foundational organizing principles in physics; however, an analogous framework for mixed, decohered quantum states has only begun to emerge. A central new concept is strong-to-weak spontaneous symmetry breaking (SW-SSB), a sharp transition in mixed quantum states that is invisible to any observable linear in the density matrix and that has since been predicted across a broad class of open and monitored quantum systems~\cite{lee2023, lessa2025, weinstein2025, gu2025, sala2024, zhang2025b, kim2024, chen2025, sa2025, ziereis2025, zerba2025, hauser2025, kuno2025, sarma2026}. It also provides a unifying language for phenomena as disparate as the decodability of topological quantum memories~\cite{lee2023, fan2024, bao2023} and the emergence of classical hydrodynamics from decohered quantum dynamics~\cite{huang2025, hauser2026}. Here we report the first experimental observation of SW-SSB, in dephased single-component fermionic matter imaged by a quantum gas microscope. A quantum-classical estimator built on a machine-learned Gaussian reference state gives direct access to the nonlinear Rényi-1 and Rényi-2 correlators that diagnose SW-SSB, and reveals long-range Rényi order in the dephased Fermi liquid. Adding a commensurate superlattice drives the underlying fermions through a metal-to-insulator transition that, after full dephasing, manifests as a sharp SW-SSB phase transition. Our results uncover the symmetry principle behind information-theoretic transitions in open quantum systems, and extend Landau's symmetry paradigm into the regime of real, decohering quantum devices.
\end{abstract}

\maketitle
%TC:endignore

% =============================================================
% Main text
% =============================================================

% \
% main.tex  --  Main text body
%
% This file is \input by manuscript.tex and contains only section
% content (no \documentclass, \begin{document}, or title block).
%
% Sections:
%   1. Introduction
%   2. SW-SSB as a new type of ODLRO
%   3. Experiments
%      3.1 Protocol
%      3.2 Dephased Fermi liquid
%      3.3 SW-SSB phase transition
% =============================================================

\section{Introduction}

Symmetry and spontaneous symmetry breaking are central organizing principles for characterizing and classifying equilibrium  states of matter. Building on these principles, Landau's classic paradigm provides an extraordinarily successful framework for describing equilibrium and zero-temperature phases of matter. In this context, the foundational idea of \textit{off-diagonal long-range order} (ODLRO) plays a special role in characterizing symmetry-broken phases like Bose-Einstein condensates and superconductors \cite{penrose1951,penrose1956,yang1962,bloch2000a}. Although certain quantum phases, such as topologically ordered states, were once thought to lie beyond any symmetry-based classification, the emergence of generalized symmetries in recent years has significantly extended the reach of the symmetry paradigm and brought topological orders under the umbrella of spontaneous symmetry breaking~\cite{nussinov2009,gaiotto2015,ji2020,choi2022,Cordova2022,cordova2023,mcgreevy2023}.

Despite these impressive advances, the landscape of \textit{mixed, out-of-equilibrium states of matter} remains largely unexplored. What has already become clear is that in order to characterize nonequilibrium states, the concepts of symmetry and symmetry breaking must again be broadened. Recently, a new pattern of symmetry has been uncovered in the context of quantum mixed states: every symmetry can have a ``\textit{strong}" and a ``\textit{weak}" version~\cite{buca2012,degroot2022}. Strongly-symmetric states have a definite symmetry-charge quantum number, while weakly-symmetric states are incoherent mixtures of states with different charge quantum numbers. As a system gradually loses its quantum features under decoherence, there can be a spontaneous breaking from strong to weak symmetry (referred to as ``SW-SSB" hereafter)~\cite{lee2023, lessa2025,weinstein2025,gu2025,sala2024, huang2025, zhang2025b, kim2024, chen2025, sa2025, ziereis2025, zerba2025, hauser2025, kuno2025, sarma2026}. SW-SSB holds both fundamental conceptual significance and direct relevance to quantum technologies: it is linked to the ``decodability transition" of topological quantum memories~\cite{lee2023,fan2024,bao2023}, and it has been identified with the emergence of classical hydrodynamics in decohered quantum systems~\cite{huang2025,hauser2026}.

In this work, we report the first experimental observation of both the SW-SSB phase and phase transition in fully-dephased fermionic states with a fixed $U(1)$ charge. This observation marks an important step toward a systematic understanding of nonequilibrium mixed quantum states. Our work also shows how the language of condensed matter physics, i.e. ODLRO and superconductivity, provides a natural framework for characterizing the information output of quantum simulators and quantum computers. In the following, we will first give a brief overview of the key quantities needed to understand SW-SSB. A more comprehensive overview is presented in the supplementary material (Sec.~\ref{sec:overview}). 

\begin{figure}[hbt]
    \centering
    \includegraphics[width=\columnwidth]{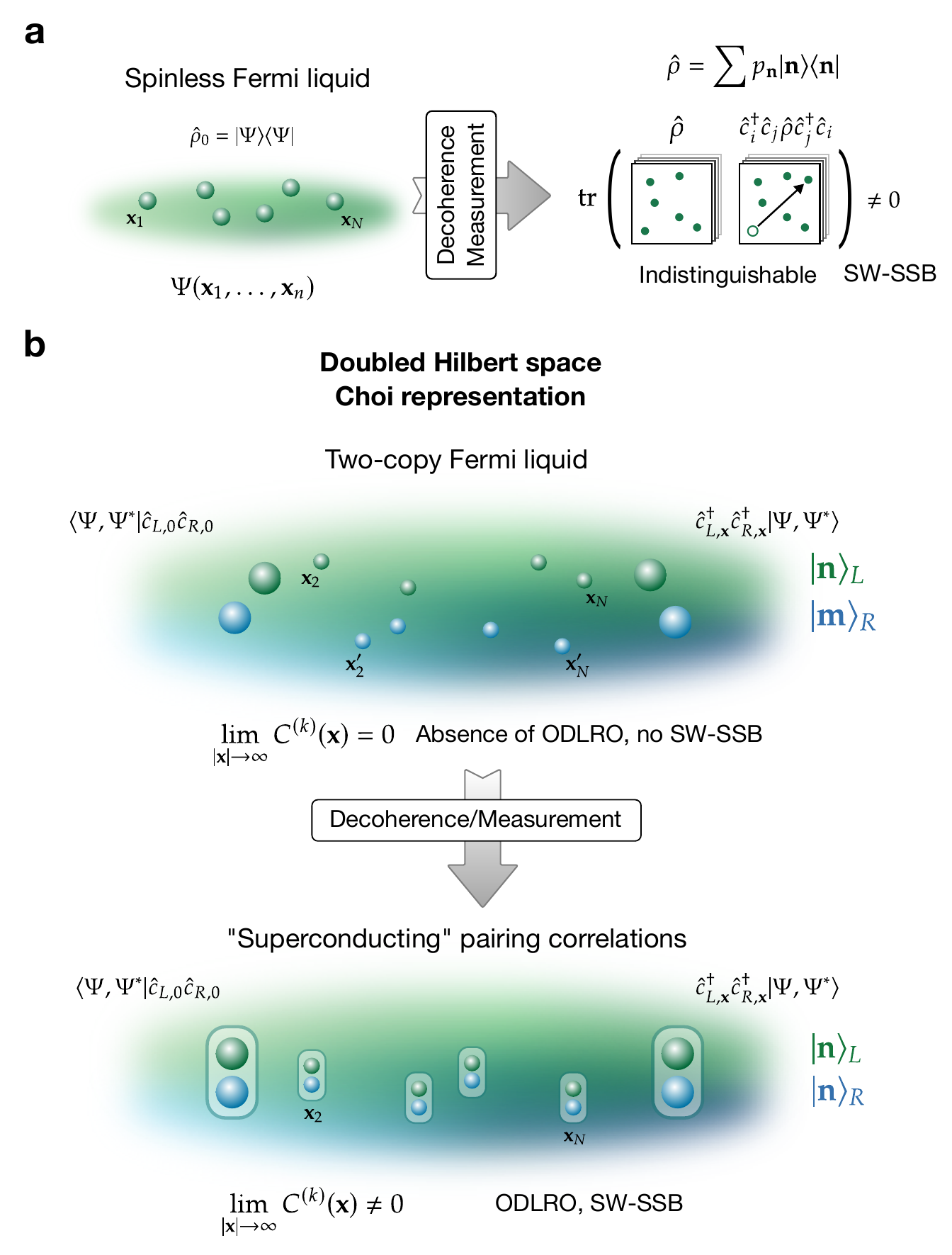}
    \caption{\textbf{Strong-to-weak spontaneous symmetry breaking as off-diagonal long-range order.}
    \textbf{(a)} The fully dephased density matrix of a quantum state prepared in an optical lattice is diagonal in the occupation basis and can be constructed from projective measurements. The R\'{e}nyi correlators measure the overlap between the resulting probability distribution and a ``displaced'' distribution, in which one particle has been moved from the origin to position $\vect x$ (Eq.~(\ref{BC})). If this overlap remains large at long distances, the two distributions are indistinguishable and the system exhibits SW-SSB.
    \textbf{(b)} In the Choi (doubled Hilbert space) representation, a spinless Fermi liquid maps to a two-component (replica) Fermi liquid. Dephasing acts as an attractive interaction between the two components, driving Cooper pair formation in the strong dephasing limit. The resulting inter-replica superconducting correlations are precisely the R\'{e}nyi-2 correlator (Eq.~(\ref{SC})), connecting SW-SSB to ODLRO in the doubled space.}
    \label{fig:fig1-swssb-schematic}
\end{figure}

\section{SW-SSB as a new type of ODLRO}

Ordinary spontaneous symmetry breaking of a U(1) symmetry in a quantum state $|\Psi_0\rangle$ is diagnosed by the presence of off-diagonal long-range order (ODLRO) \cite{penrose1951,penrose1956,yang1962,bloch2000a}:
\beq
\lim_{|\vect{x}| \ra \infty} \langle \Psi_0 | \hat{b}_0 \, \hat{b}^\dagger_\vect{x} | \Psi_0 \rangle
= \tr\left( \hat \rho_0 \, \hat{b}_0 \, \hat{b}^\dagger_\vect{x} \right) \neq 0,
\eeq
where $\hat{b}^\dagger_\vect{x}$ ($\hat{b}_\vect{x}$) denotes a boson creation (annihilation) operator at position $\vect{x}$ and $\hat \rho_0$ is the pure state density matrix of the system. SW-SSB is hidden in a ``finer structure" of the density matrix, and diagnosing it is more involved. In fact, it is not possible to observe SW-SSB in quantities linear in the density matrix. Instead, U(1) SW-SSB can be diagnosed by the following \textit{R\'{e}nyi-$k$ correlators}, which are nonlinear in the density matrix:
\beq
C^{(k)}(\vect{x}) = \frac{\tr\left( \hat \rho^{k/2} \, \hat{c}_0 \, \hat{c}^\dagger_\vect{x} \, \hat \rho^{k/2} \,
\hat{c}_0^\dagger \, \hat{c}_\vect{x} \right)}{\tr \left( \hat \rho^k \right) },
\eeq
where $\hat{c}^\dagger_\vect{x}$ is a particle creation operator (here the particles may be fermions or bosons, but henceforth we take them as fermions). The R\'{e}nyi-$1$ correlator, $C^{(1)}$, is tied to the precise classification of mixed state phases of matter~\cite{lessa2025,weinstein2025}. The R\'{e}nyi-$2$ correlator, $C^{(2)}$, is the simplest replica generalization and is more convenient to evaluate theoretically. In this work, we probe both for the case of fully dephased quantum matter.

\medskip

\textit{Distinguishability of probability distributions.} ---
Locally-resolved measurements of particle coordinates, i.e. ``snapshot" measurements, can be treated as an infinite-strength dephasing channel that drives an initial state $\hat \rho_0$ to a diagonal density matrix $\hat\rho$ given by
\beq
\hat \rho = \sum_{\textbf{n}} p_{\mathbf{n}}
|\bf n\rangle \langle \bf n |,
\eeq
where $\bf{n}$ stands for a bit string configuration of site occupancies $n_i=0,1$ and $p_{\bf n}=\langle\bf{n}|\hat\rho_0|\bf{n}\rangle$ is the probability of measuring $\bf{n}$. The R\'{e}nyi correlators of the dephased state $\hat\rho$ can be written directly in terms of $p_{\bf n}$: 
\beq
\begin{aligned}
C^{(2)}(\vect{x}) &= \frac{\sum_{\bf n} p_{\bf n} p_{\bf n'} }{\sum_{\bf n} p^2_{\bf n}}, \\
C^{(1)}(\vect{x}) &= 
\sum_{\bf n} \sqrt{p_{\bf n}} \sqrt{p_{\bf n'}}, \label{BC}
\end{aligned}
\eeq
where $\bf n'$ is the displaced configuration obtained by moving one particle from the origin to position $\vect{x}$. Both correlation functions have a simple physical interpretation: they measure the overlap between the original probability distribution $p_{\bf n}$ and the displaced distribution $p_{\bf n'}$ (Fig.~\ref{fig:fig1-swssb-schematic}(a)). The R\'{e}nyi-1 correlator is identical to the Bhattacharyya Coefficient~\cite{bhattacharyya1943,hauser2026}, a standard measure of statistical similarity. 
If $C^{(k)}(\vect x)$ decays as $\sim\exp(-|\vect x|/\xi^{(k)})$, then the distributions are distinguishable on length scales larger than $\xi^{(k)}$.
If $C^{(k)}(\vect x)$ remains large even when a particle is displaced over large distances $|\vect{x}|\rightarrow \infty$, the two distributions are fundamentally 
indistinguishable: the occupation statistics of the system are insensitive to moving a single particle. This is the signature of SW-SSB~\cite{hauser2026}. As an example, a canonical-ensemble Gibbs state at infinite temperature exhibits maximal indistinguishability: all density configurations have equal probability, and the R\'{e}nyi correlators $C^{(k)}(\vect x)=\bar n(1-\bar n)$ are a constant for filling fraction $\bar n$. Generally, the R\'{e}nyi correlators increase with temperature, as higher temperatures lead to a more uniform distribution $p_{\bf n}$ that is more self-similar under displacement of single particles.

\textit{Choi representation and Cooper pair correlations.} ---
Further insight can be gained through the ``doubled state" (Choi) representation, which maps any density matrix to a pure quantum state in a doubled Hilbert space~\cite{jamiolkowski1972,choi1975,su2025}:
\beqn
\hat{\rho} = \sum_{\alpha\beta}\rho_{\alpha\beta} |\alpha\rangle\langle\beta | \  \ra \  |\rho \rrangle = \sum_{\alpha\beta} \rho_{\alpha\beta} |\alpha\rangle_L \otimes |\beta\rangle_R.
\eeqn
The ``Left" (L) and ``Right" (R) replica spaces represent the ket and bra spaces of the density matrix. Defining the inter-replica pairing operator between these Left and Right spaces as $\hat{\Delta}_\vect{x} = \hat{c}_{L,\vect{x}} \, \hat{c}_{R,\vect{x}}$, the R\'{e}nyi-2 correlator maps exactly to the inter-replica Cooper pair correlation function:
\beqn 
C^{(2)}(\vect{x}) = \frac{\tr\left( \hat \rho \, \hat{c}_0 \, \hat{c}^\dagger_\vect{x} \, \hat \rho \, \hat{c}_\vect{x} \, \hat{c}_0^\dagger \right)}{\tr \hat \rho^2} = \frac{\llangle \rho | \hat{\Delta}_0 \, \hat{\Delta}^\dagger_\vect{x} |\rho\rrangle}{\llangle \rho | \rho \rrangle}. \label{SC} \eeqn
In other words, the R\'{e}nyi-2 correlator is nothing but the inter-replica Cooper pair correlation function in the doubled space, and SW-SSB corresponds to ODLRO of Cooper pairs in the Choi representation (see Sec.~\ref{sec:overview}).

In the doubled space, density dephasing acts as an attractive density-density interaction between the Left and Right spaces:
\beqn |\rho(g) \rrangle \sim e^{ - \sum_i g (\hat{n}_{L,i} - \hat{n}_{R,i})^2} | \rho_0\rrangle. \eeqn
In the strong dephasing limit ($g \rightarrow \infty$), the coordinates of particles from the two spaces are tightly bound together (Fig.~\ref{fig:fig1-swssb-schematic}(b)).  When dephasing is introduced via locally-resolved measurements, the pair size is given by the resolution of the detection system. If the initial quantum state $\hat\rho_0$ is a spinless Fermi liquid, the doubled state $|\rho_0\rrangle$ is a two-component Fermi liquid. After full dephasing, the fermions form tightly-bound Cooper pairs in the doubled space, leading to superconducting correlations. Superconductivity in the doubled space maps precisely onto SW-SSB, as shown in Eq.~(\ref{SC}) and illustrated in Fig.~\ref{fig:fig1-swssb-schematic}(b) (see Sec.~\ref{sec:supmat:choi}). By contrast, if the initial state is a crystalline insulator, the effective attractive interaction simply produces a bosonic crystalline insulator in the doubled space -- thus, there is no SW-SSB.

\textit{Off-diagonal long-range order.} ---
More generally, the R\'{e}nyi-$k$ correlators can be expressed as standard ODLRO,
\beq
C^{(k)}(\vect{x})
\sim \int \prod_{i=2}^{N} d^dx_i \;
\Phi_k(0, \vect{x}_2, \cdots \vect{x}_N) \Phi_k(\vect{x}, \vect{x}_2, \cdots \vect{x}_N),
\eeq
if we view $p^{k/2}({\bf n})$ as a first-quantized bosonic wave function $\Phi_k(\{\vect{x}_i\})$. If $\hat \rho_0$ is a pure-state wavefunction, $\Psi(\{\vect{x}_i\})$, then the R\'{e}nyi-2 formulation is, after dephasing, mathematically equivalent to superconductivity: the amplitude $\Phi_2(\{\vect{x}_i\})=|\Psi(\{\vect{x}_i\})|^2$ is just that of a Cooper pair. The R\'{e}nyi-1 formulation instead detects ODLRO in the modulus of the wavefunction, $\Phi_1(\{\vect{x}_i\})=|\Psi(\{\vect{x}_i\})|$, which can be interpreted as a non-negative wavefunction of bosons. Intuitively, as long as the original state $\hat\rho_0$ is in the quantum degenerate regime, i.e. the typical wavelength of particles is greater than the interparticle distance, the dephased boson wave functions $\Phi_k(\{\vect{x}_i\})$ are expected to exhibit ODLRO; hence, there is SW-SSB. Examples of such highly degenerate quantum states of fermions include the metallic Fermi liquid, integer and fractional quantum Hall states~\cite{sarma2025,wang2025a,kiely2025}.

\section{Experiments}

\subsection{Protocol}

In this work we study SW-SSB in fully-dephased spinless fermionic states. We probe these states by producing a weakly-interacting spin-$1/2$ lattice Fermi gas in a quantum gas microscope. We then perform site-resolved fluorescence measurements of all fermion positions simultaneously, but only for a single spin species. The result is a collection of ``snapshots" $\bf n$ of the initial spinless state $\hat\rho_0$, which are sampled with probability $p_{\bf n}=\langle\bf n|\hat\rho_0|\bf n\rangle$.

As the fully-dephased density matrix of spinless fermions is diagonal in the spatial occupation basis, the R\'{e}nyi correlators can be constructed directly from these snapshots. In principle, with enough measurements, one could construct the full normalized probability distribution function (and thereby extract $C^{(k)}(\vect x)$), but this would require an exponentially large number of snapshots. Instead, we employ a ``quantum-classical" (QC) cross correlator as a proxy for the  R\'{e}nyi correlators~\cite{garratt2023,zhang2025probing}. That is, we define

\beq
\begin{aligned}
C_{\rm QC}^{(2)}(\vect x) &= \frac{\sum_{\bf n} p_{\bf n}p^c_{\bf n'}}{\sum_{\bf n}p_{\bf n}p^c_{\bf n}}\approx\frac{\sum_{m} p^c_{{\bf n}_m'}}{\sum_{m}p^c_{{\bf n}_m}}, \\
C_{\rm QC}^{(1)}(\vect x) &= \sum_{\bf n} p_{\bf n}  \sqrt{ \frac{p^c_{\bf n'}}{p^c_{\bf n}} }\approx \frac{1}{M}\sum_{m}\sqrt{ \frac{p^c_{{\bf n}_m'}}{p^c_{{\bf n}_m}} }, \label{eq:QC}
\end{aligned}
\eeq
where $p^c_{\bf n}$ is an occupation probability distribution generated by a classical computer.
In the second equality, we approximate $C^{(k)}_{\rm QC}$ by noting that the $M$ experimental snapshots ${\bf n}_m$ are sampled directly from $p_{\bf n}$. These expressions serve as estimators for the true correlation functions, $C^{(k)}$, provided the classical model faithfully reproduces the experimental distribution.

As the experimental system is weakly-interacting, the classical model we employ is a Gaussian (non-interacting) fermionic state. The static properties of Gaussian fermionic states, including configuration probabilities $p^c_{\bf n}$, are determined solely by their equal-time Green's function, $G_{ij}=\langle\hat c^\dagger_i\hat c_j\rangle$. The particular Gaussian state we use is determined by choosing $G_{ij}$ to minimize the Kullback--Leibler (KL) divergence between $p_{\bf n}$ and $p^c_{\bf n}$ (see Sec.~\ref{sec:suppmat:theory}). This procedure produces the ``closest" Gaussian state to the experimental system, given the set of snapshot measurements ${\bf n}_m$. To ensure the reliability of the QC estimators (Eq.~\eqref{eq:QC}), we compare them with ``classical-classical" (CC) correlators, which are exact R\'{e}nyi correlators of the classical distribution $p^c_{\bf n}$. If the correlators match, it indicates that the QC estimator is a good approximation for the experimental system.

\begin{figure}[ht]
    \centering
    \includegraphics[width=\linewidth]{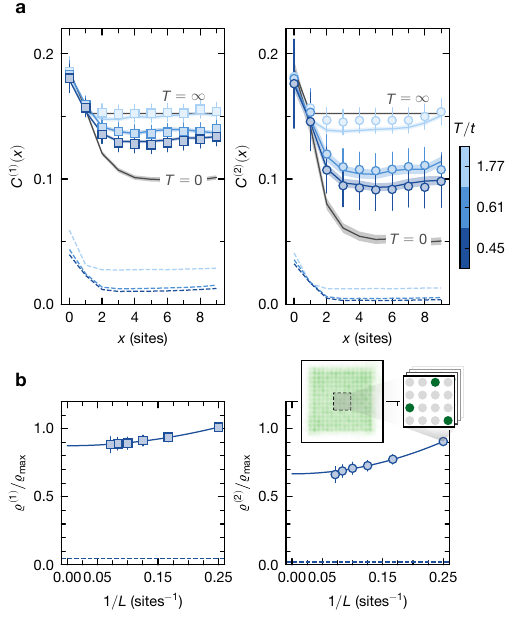}
    \caption{\textbf{Dephased Fermi liquid.}
    \textbf{(a)}~The R\'{e}nyi-1/Fidelity and R\'{e}nyi-2 correlators of the fully-dephased Fermi liquid state, based on snapshots extracted from the experiment at various temperatures. 
    The quantum-classical estimators (Eq.~\eqref{eq:QC})
    are shown as squares /circles,
    where the classical state is a machine-learned Gaussian fermionic state trained solely on the experimental snapshots. The corresponding classical-classical %(CC) 
    correlator is shown by the solid lines.
    At high temperature, the dephased R\'{e}nyi correlators approach the infinite-temperature value $\bar{n} ( 1 - \bar{n}) \sim 0.15$.
    Dashed lines show the R\'{e}nyi correlators of the undephased Fermi liquid state,  
    computed using the same machine-learned classical
    states. 
    \textbf{(b)}~Quantum-classical R\'{e}nyi-$k$ condensate densities (Eq.~(\ref{chi})) of $L\times L$-site subsystems versus $1/L$ at the lowest experimental temperature $T/t=0.45$ (squares/circles). Condensate densities are normalized by the maximum condensate density $\varrho_{\rm max}=\bar n(1-\bar n)$, so we refer to them as condensate fractions. We see that both the R\'{e}nyi-1 and 2 condensate fractions converge to a constant as $L\to\infty$, as shown by the least-squares fit (solid curve). The condensate fraction in the undephased system (dashed line) is more than an order of magnitude smaller. In \textbf{(a)} and \textbf{(b)}, error bars for squares/circles are smaller than the marker size where not visible; shaded regions for solid lines denote the uncertainty and are thinner than the line width where not visible. Results for updephased Fermi liquid states are exact.
    }
    \label{fig:fig2-renyicorrelator}
\end{figure}

\subsection{Dephased Fermi liquid}

We first consider the case of a dephased 2d Fermi liquid, which is expected to exhibit SW-SSB~\cite{sarma2026}. Our experiments begin by loading a degenerate Fermi gas of $^6$Li atoms in a balanced mixture of two internal spin states into a square optical lattice, producing a homogeneous, square-shaped system of approximately 196 lattice sites. The lattice depths are chosen such that the tunneling amplitude is much smaller than the bandgap, and the inter-spin interaction is tuned to a negligible value. The quantum gas microscope setup allows for site-resolved resolution \cite{gross2017,gross2021}. 
We emphasize that the spin of $^6$Li atoms should not be confused with the two-component structure arising in the doubled-space representation, which is an effective degree of freedom: even single-component fermions map onto two-component fermions in the Choi representation. In the experiment, we therefore detect only one spin component of the mixture, effectively realizing a single-component Fermi gas, which at low density is well described by a non-interacting tight-binding Hamiltonian,
\beq
\hat{H} = -t\sum_{\langle ij \rangle}
\left( \hat{c}^\dagger_i \hat{c}_j + \text{H.c.} \right),
\label{eq:hoppingH}
\eeq
where $\langle ij \rangle$ denotes nearest-neighbor (NN) sites, $\hat{c}^\dagger_i$ is the fermionic creation operator at site $i$, and $t$ is the NN tunneling amplitude.

In Fig.~\ref{fig:fig2-renyicorrelator}(a) we plot the QC estimators (squares/circles) and CC correlators (solid lines) of the dephased Fermi liquid density matrix at filling fraction $\bar n=0.1875$ and for various experimental temperatures. For details on thermometry, see Sec.~\ref{sec:thermometry}. While the entire system spans $14\times14$ sites, here we consider the central $12\times12$-site subregion and average the correlation functions over all sites $i$ and $j$ separated by $(x, 0)$. The close agreement between the QC and CC correlators indicates that the machine-learning algorithm has produced a  Gaussian approximation that is in good agreement with the experimental system. Both $C^{(1)}$ and $C^{(2)}$ converge to a finite value at long distances, confirming long-range order. As expected, the asymptotic value of both correlators increases with temperature. 

We compare these results for the dephased Fermi liquid to those of the undephased Fermi liquid (dashed lines), which can be computed exactly (see Sec.~\ref{sec:suppmat:theory}). As noted earlier, finite temperature induces long-range order even in the absence of dephasing. This effect is quite small, though, which clearly demonstrates that dephasing is essential for the emergence of SW-SSB correlations.

To quantify the role of finite-size effects, we define the R\'{e}nyi-$k$ condensate density as
\beqn 
\varrho^{(k)} = \frac{1}{N_{\textrm {site}}^2}\sum_{i,j} C^{(k)}(i, j),\label{chi}
\eeqn
where $N_{\textrm{site}}=L^2$ is the total number of sites. $\varrho^{(k)}$ measures the strength of the SW-SSB phase and should converge to a constant in the thermodynamic limit $L \ra \infty$. 
Considering $L\times L$-site central regions of varying sizes $L$, we measure the QC R\'{e}nyi-$k$ condensate density at $T/t=0.45$ and plot the results versus $1/L$ in Fig.~\ref{fig:fig2-renyicorrelator}(b). We normalize the condensate density by its maximal value, $\varrho_{\rm max}=\bar n(1-\bar n)$, thereby providing a definitive scale for the strength of SW-SSB. We indeed find that the condensate fraction approaches a constant as $1/L\to 0$, confirming SW-SSB in the dephased Fermi liquid. The undephased Fermi liquid, by contrast, shows weak correlations that are an order of magnitude smaller.

\begin{figure*}[ht]
    \centering
    \includegraphics[width=\textwidth]{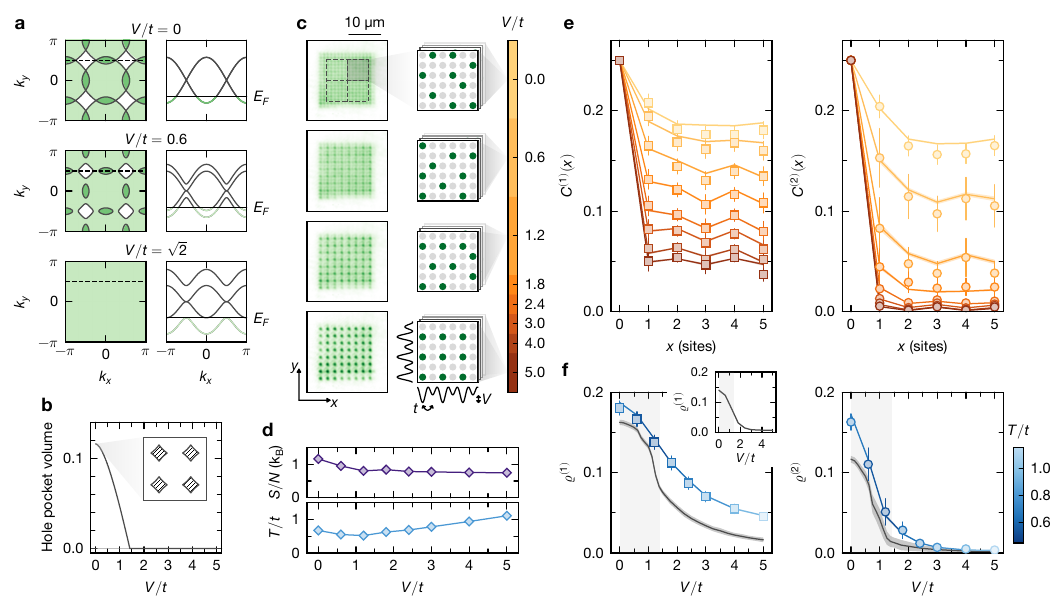}
    \caption{\textbf{SW-SSB phase transition.} \textbf{(a)}~Evolution of Fermi surface of Eq.~\eqref{eq:hoppingHdelta} while increasing $V/t$. Dark and white regions correspond to electron and hole pockets. The Fermi surface vanishes at critical $V_c/t = \sqrt{2}$. \textbf{(b)}~Size of the Fermi surface versus $V/t$, quantified by the fractional volume of the hole pockets in the Brillouin zone. \textbf{(c)}~Left: Averaged site-resolved fluorescence pictures of single-spin atoms for different $V/t$. Right: For the analysis of Rényi correlators, we use $6\times6$-site snippets cropped from the full system and postselect on configurations with a filling fraction of 25\%. 
    At low $V/t$, metallic states show strong shot-to-shot occupation fluctuations that average to a spatially uniform fluorescence profile. 
    At high $V/t$, insulating states display reproducible occupation patterns in each snapshot, closely matching the averaged fluorescence distribution. The superlattice drawing is schematic and not to scale.
    \textbf{(d)}~Entropy per particle and temperature evaluated using postselected snapshots of 
    $6\times 6$ snippets at 25\% filling (see Sec.~\ref{sec:thermometry}). Solid lines connect the data points and serve as guides to the eye; error bars are smaller than the marker size. \textbf{(e)} Quantum-classical (squares/circles) and classical-classical (solid lines) R\'{e}nyi-$k$ correlators for various $V/t$.
    As $V/t$ increases, the system transitions from a dephased metal to a dephased band insulator. Correspondingly, the long-range Rényi correlations of the dephased state progressively vanish. \textbf{(f)} Quantum-classical (squares/circles) and classical-classical (solid colored lines) R\'{e}nyi-$k$ condensate densities versus $V/t$. Colors indicate the experimental temperatures at different $V/t$ (see colorbar). Dark solid curves show classical-classical R\'{e}nyi condensate density on $6\times6$ snippets of a $14\times14$-site lattice, computed using Eq.~\eqref{eq:hoppingHdelta} at zero temperature. {\bf (inset)} Classical-classical R\'{e}nyi-1 condensate density on a $12\times12$-site subsystem of a $14\times14$-site lattice, also computed using Eq.~\eqref{eq:hoppingHdelta} at zero temperature. Note the sharper change across $V_c/t$ relative to the $6\times 6$ data. In \textbf{(e)} and \textbf{(f)}, error bars for squares/circles are smaller than the marker size where not visible; shaded regions for solid lines denote the uncertainty and are thinner than the line width where not visible.}
    \label{fig:fig3-swssb}
\end{figure*}

\subsection{SW-SSB phase transition}

It is expected that a metal-to-insulator transition in the underlying unitary problem translates, after full dephasing, into a SW-SSB phase transition in the mixed ensemble. To check this, we employ a superlattice to generate a square lattice with a site-dependent potential and thus realize a modified version of the Hamiltonian in Eq.~\eqref{eq:hoppingH}, defined up to a global energy offset as
\beq
\hat{H}
= -t\sum_{\langle ij \rangle}
\left(
\hat{c}_i^\dagger \hat{c}_j
+ \text{H.c.}
\right)
+ \sum_i \delta_i \hat{n}_i,
\label{eq:hoppingHdelta}
\eeq
where $\hat{n}_i$ is the atom number operator at site $i$, and the on-site potential $\delta_i$ follows a four-site periodic pattern $(0, V, V, 2V)$ within each unit cell, which can be written equivalently as $\delta(x,y)=V - \frac{V}{2} \left[(-1)^x + (-1)^y \right]$ (for integer site labels, $x,y$). We choose the density of fermions to be commensurate, with one (spinless) fermion for each super-lattice unit cell. For large $V/t$, the ground state is a spinless band insulator; for small $V/t$, it is a metal. As illustrated in Fig.~\ref{fig:fig3-swssb}(a, b), they are separated by a Lifshitz transition at $V_c/t=\sqrt{2}$ where the Fermi surface vanishes.

In Fig.~\ref{fig:fig3-swssb}(c) we show the average experimental density profile for various $V/t$, illustrating the transition from a uniform metal to a localized charge crystal. The experimental states are formed by ramping down $V/t$ from a large initial value (see Sec.~\ref{sec:experimental_sequence}). Panel (d) shows the entropy densities $S/N$ and the temperatures $T/t$ for various $V/t$ computed from the experimental snapshots and the Hamiltonian in Eq.~\eqref{eq:hoppingHdelta} (see Sec.~\ref{sec:thermometry}). We achieve nearly constant entropy density $S/N\sim0.8k_{\rm B}$ down to the critical $V_c/t$.

In Fig.~\ref{fig:fig3-swssb}(e) we plot the QC correlators (squares/circles) and CC correlators (solid lines) for various $V/t$, which again agree remarkably well with one another. The correlation functions are computed by partitioning the full optical lattice into four $6\times 6$-site regions, which allows us to obtain $\sim1000$ snapshots at the desired filling fraction. Averaging over distance, we see a strong suppression of the R\'{e}nyi correlation as $V/t$ is increased. In Fig.~\ref{fig:fig3-swssb}(f) we compute the associated QC and CC condensate densities (squares/circles and lines, respectively), as well as a zero-temperature CC curve (gray) computed with respect to the Hamiltonian in Eq.~\eqref{eq:hoppingHdelta}. We find that $\varrho^{(2)}$ vanishes sharply as we tune $V/t$ through the critical point, while $\varrho^{(1)}$ changes more gradually. This distinction is largely explained by the expectation that $\varrho^{(2)} \sim \left( \varrho^{(1)} \right)^2$ for large $V/t$, which is discussed in Sec.~\ref{sec:suppmat:theory}.
The relative smoothness of the QC cross correlators may otherwise be attributed to finite size and entropy density. 

Our results strongly support the existence of a SW-SSB transition driven by superlattice potential $V/t$.

\textit{Outlook.} --- Our results open several avenues for future investigation. On the theoretical side, determining the universality class of the SW-SSB phase transition observed here is an outstanding challenge: heuristically, in the doubled space, it corresponds to a gauged superconductor-insulator transition, connecting to long-standing questions about deconfined criticality and spin liquid physics. On the experimental side, extending this work to interacting fermionic systems, such as the Fermi-Hubbard model, would probe SW-SSB in strongly correlated matter. Extending our work to Dirac fermions rather than Fermi liquids would allow us to probe Choi-spin liquids with emergent gauge structure~\cite{su2025}. Application of our measurement protocol to some topological quantum states is predicted to exhibit quasi-long-range SW-SSB upon dephasing~\cite{sarma2025,wang2025a,kiely2025}. Going beyond static dephasing, time-resolved measurements of R\'{e}nyi correlators during controlled decoherence could directly track the dynamical buildup of SW-SSB and test recent predictions about the emergence of classical hydrodynamics~\cite{hauser2026}. More broadly, the framework developed here suggests that R\'{e}nyi correlators and SW-SSB diagnostics could serve as a general tool for characterizing the information output of noisy quantum devices, complementing existing benchmarks based on fidelity and entanglement measures. 

Finally, we note that our evaluation of the Rényi correlator relies only on ``learning" and reconstructing the density distribution of the experimental system, which is less demanding than a full simulation of the quantum state because it requires only information about the modulus of the wave function. We will explore the full potential of this approach in future work.
\medskip

{\bf Acknowledgments --- }
We thank Andreas von Haaren for experimental support. We acknowledge helpful discussions with Yizhi You. T.G.K. acknowledges support from the Gordon and Betty Moore Foundation under GBMF7392 and from the National Science Foundation under grant PHY-2309135 to the Kavli Institute for Theoretical Physics (KITP). CX is supported by the Simons foundation through the Simons Investigator program. This work has also been supported by the Simons Collaboration on Ultra-Quantum Matter, which is a grant from the Simons Foundation (651457, M.P.A.F.). M.P.A.F. is  also supported by a Quantum Interactive Dynamics grant from the William M. Keck Foundation. Further support was provided by the Max Planck Society (MPG), the Horizon Europe program HORIZON-CL4-2022-QUANTUM-02-SGA (project 101113690, PASQuans2.1), the German Federal Ministry of Research, Technology and Space (BMFTR grant agreement 13N15890, FermiQP), and Germany’s Excellence Strategy (EXC-2111-390814868). P.M.P. acknowledges funding from the European Union’s Horizon 2020 research and innovation programme (Grant Agreement No. 948240 — ERC Starting Grant UniRand).

% =============================================================
% Bibliography
% =============================================================

\bibliography{references}

% =============================================================
% Supplementary Information
% =============================================================
%TC:ignore
% =============================================================
% supplements.tex  --  Supplementary Information
%
% This file is \input by manuscript.tex after the bibliography.
% Counters are reset so that equations, figures, and tables
% are numbered S1, S2, etc.
%
% Sections:
%   S1. Experimental details
%   S2. Theory section
%   S3. Thermometry
%   S4. Overview of SW-SSB
% =============================================================

% Reset counters for supplementary numbering (S1, S2, ...)
\setcounter{section}{0}
\setcounter{equation}{0}
\setcounter{figure}{0}
\setcounter{table}{0}

% Display format: S1, S2, ...
\renewcommand{\theequation}{S\arabic{equation}}
\renewcommand{\thefigure}{S\arabic{figure}}
\renewcommand{\thetable}{S\arabic{table}}
\renewcommand{\thesection}{S\arabic{section}}

% Unique hyperref anchors to avoid duplicate-destination warnings
\renewcommand{\theHequation}{Sequation.\arabic{equation}}
\renewcommand{\theHfigure}{Sfigure.\arabic{figure}}
\renewcommand{\theHtable}{Stable.\arabic{table}}
\renewcommand{\theHsection}{Ssection.\arabic{section}}
\renewcommand{\v}[1]{{\bf #1}}
\newcommand{\sign}{{\rm sign}}
\newcommand{\M}{{\cal{M}}}
\def\vol{{\rm Vol}}
\newcommand{\psib}{{\bar{\psi}}}
\newcommand{\rb}{{\bar{\rho}}}
\newcommand{\phib}{{\bar{\phi}}}
\newcommand{\dt}{{\Delta}}
\newcommand{\w}{{\omega}}
\newcommand{\zh}{{\hat{z}}}
\newcommand{\qh}{{\hat{q}}}
\newcommand{\gr}{{\nabla}}
\newcommand{\p}{\partial}
\newcommand{\pt}{\partial_t}
\newcommand{\sgn}{{\rm sgn}}
\renewcommand{\t}[1]{{\tilde #1}}
\newcommand{\ua}{\uparrow}
\newcommand{\da}{\downarrow}
\newcommand{\etc}{{\it etc~}}
\newcommand{\etal}{{\it et al.~}}
\newcommand{\veps}{\varepsilon}
\newcommand{\vphi}{\varphi}
\newcommand{\cA}{ {\cal A} }
\newcommand{\cB}{ {\cal B} }
\newcommand{\cC}{ {\cal C} }
\newcommand{\cD}{ {\cal D} }
\newcommand{\cE}{ {\cal E} }
\newcommand{\cF}{ {\cal F} }
\newcommand{\cG}{ {\cal G} }
\newcommand{\cH}{ {\cal H} }
\newcommand{\cK}{ {\cal K} }
\newcommand{\cL}{ {\cal L} }
\newcommand{\cP}{ {\cal P} }
\newcommand{\cT}{ {\cal T} }
\newcommand{\cS}{ {\cal S} }
\newcommand{\cV}{ {\cal V} }
\newcommand{\cZ}{ {\cal Z} }
\newcommand{\Int}{\mathrm{int}}
\newcommand{\ii}{\mathrm{i}}
\newcommand{\bx}{\mathbf{x}}
\newcommand{\by}{\mathbf{y}}
\newcommand{\br}{\mathbf{r}}
\newcommand{\bo}{\mathbf{0}}
\newcommand{\hrho}{\hat{\rho}}
\newcommand{\htheta}{\hat{\theta}}
\newcommand{\hphi}{\hat{\phi}}
\newcommand{\hn}{\hat{n}}
\newcommand{\hO}{\hat{O}}
\newcommand{\hA}{\hat{A}}
\newcommand{\hvA}{\hat{\vect{A}}}
\newcommand{\hB}{\hat{B}}
\newcommand{\hvB}{\hat{\vect{B}}}
\newcommand{\hE}{\hat{E}}
\newcommand{\vE}{\vect{E}}
\newcommand{\hP}{\hat{P}}
\newcommand{\hvE}{\hat\vect{{E}}}
\newcommand{\hh}{\hat{h}}
\newcommand{\vh}{\vect{h}}
\newcommand{\hvh}{\hat{\vect{h}}}
\newcommand{\vectr}{\vect{r}}
\newcommand{\bvectr}{\bar{\vect{r}}}
\newcommand{\vectl}{\vect{l}}
\newcommand{\hV}{\hat{V}}
\newcommand{\SO}{\mathrm{SO}}
\renewcommand{\O}{\mathrm{O}}
\newcommand{\SU}{\mathrm{SU}}
\newcommand{\U}{\mathrm{U}}
\newcommand{\Sp}{\mathrm{Sp}}
\newcommand{\Gr}{\mathrm{Gr}}
\newcommand{\bbC}{\mathbb{C}}
\newcommand{\bbD}{\mathbb{D}}
\newcommand{\bbE}{\mathbb{E}}
\newcommand{\bbH}{\mathbb{H}}
\newcommand{\bbI}{\mathbb{I}}
\newcommand{\bbN}{\mathbb{N}}
\newcommand{\bbP}{\mathbb{P}}
\newcommand{\bbQ}{\mathbb{Q}}
\newcommand{\bbR}{\mathbb{R}}
\newcommand{\bbS}{\mathbb{S}}
\newcommand{\bbU}{\mathbb{U}}
\newcommand{\bbV}{\mathbb{V}}
\newcommand{\bbZ}{\mathbb{Z}}
\newcommand{\calA}{\mathcal{A}}
\newcommand{\calB}{\mathcal{B}}
\newcommand{\calC}{\mathcal{C}}
\newcommand{\calD}{\mathcal{D}}
\newcommand{\calE}{\mathcal{E}}
\newcommand{\calF}{\mathcal{F}}
\newcommand{\calG}{\mathcal{G}}
\newcommand{\calH}{\mathcal{H}}
\newcommand{\calI}{\mathcal{I}}
\newcommand{\calJ}{\mathcal{J}}
\newcommand{\calK}{\mathcal{K}}
\newcommand{\calL}{\mathcal{L}}
\newcommand{\calM}{\mathcal{M}}
\newcommand{\calN}{\mathcal{N}}
\newcommand{\calO}{\mathcal{O}}
\newcommand{\calP}{\mathcal{P}}
\newcommand{\calQ}{\mathcal{Q}}
\newcommand{\calR}{\mathcal{R}}
\newcommand{\calS}{\mathcal{S}}
\newcommand{\calT}{\mathcal{T}}
\newcommand{\calU}{\mathcal{U}}
\newcommand{\calV}{\mathcal{V}}
\newcommand{\calW}{\mathcal{W}}
\newcommand{\calX}{\mathcal{X}}
\newcommand{\calY}{\mathcal{Y}}
\newcommand{\calZ}{\mathcal{Z}}
%\newcommand{\tr}{\mathrm{tr}}
%\DeclareUnicodeCharacter{2215}{\ensuremath{/}}
\newcommand{\bbrakket}[2]{\mbox{$ \langle\!\langle #1 | #2 \rangle\!\rangle $}}
\newcommand{\kket}[1]{\mbox{$| #1 \rangle\!\rangle$}}
\newcommand{\bbra}[1]{\mbox{$\langle\!\langle #1 |$}}
% Force a new page and typeset the SI title manually,
% since revtex only supports a single \maketitle per document.
\clearpage
\onecolumngrid
\begin{center}
    {\Large \textbf{Supplementary Information for:} \\[0.5em]
    \Large \textbf{Observation of Strong-to-Weak Spontaneous Symmetry Breaking \\ in a Dephased Fermi Gas}}
\end{center}
\vspace{2em}
\twocolumngrid

%%%%%%%%%%%%%%%%%%%%%%%%%%%%%%%%%%%%
% EXPERIMENTAL DETAILS
%%%%%%%%%%%%%%%%%%%%%%%%%%%%%%%%%%%%

%TC:ignore

\section{Experimental details} \label{sec:experiment}

\subsection{Optical superlattice potential}
\label{sec:superlattice}
In our quantum simulator, a degenerate Fermi gas of spin-1/2 $^6$Li atoms is loaded into a two-dimensional optical superlattice generated by the interference of laser beams at 532 and 1064\,nm~\cite{chalopin2025}. The resulting potential $V_{\mathrm{tot}}(x, y)$ takes the form

\begin{equation}
\begin{split}
V_{\mathrm{tot}}(x,y) &= V_{x,\mathrm{S}} \cos^2\!\biggl(\frac{\pi x}{a_x} + \phi_x\biggr)
         - V_{x,\mathrm{L}} \cos^2\!\biggl(\frac{\pi x}{2a_x}\biggr) \\
        &\quad + V_{y,\mathrm{S}} \cos^2\!\biggl(\frac{\pi y}{a_y} + \phi_y\biggr)
         - V_{y,\mathrm{L}} \cos^2\!\biggl(\frac{\pi y}{2a_y}\biggr),
\end{split}
\end{equation}
where $a_x = \SI{1.11}{\micro\m}$ and $a_y = \SI{1.14}{\micro\m}$ are the lattice constants set by the shallow-angle lattice geometry, $V_{x,\mathrm{S}}$ and $V_{y,\mathrm{S}}$ denote the short-lattice depths, while $V_{x,\mathrm{L}}$ and $V_{y,\mathrm{L}}$ the long-lattice depths along $x$ and $y$, and $\phi_x$ and $\phi_y$ are the relative superlattice phases. Lattice depths are expressed in units of the respective recoil energies \si{\Eris} $ = h^2/8ma^2_i$ for the short lattices and \si{\Eril} $ = \si{\Eris}/4$ $(i = x, y)$ for the long lattices, where $h$ is Planck's constant and $m$ the atomic mass.

Tuning the short- and long-lattice depths together with their relative phases provides control over both the NN tunneling amplitudes $t_{x,y}$ and the inter-site detuning $V_{x,y}$ along the $x$ and $y$ directions, both obtained from the band structure of $V_{\mathrm{tot}}(x,y)$. In this work, we employ the configuration $\phi_x = \phi_y = \pi/2$, the anti-symmetric phase, in which the NN tunneling amplitudes are spatially homogeneous along each lattice direction while the inter-site detuning is maximized for a given set of lattice depths. The lattice depths are further adjusted to ensure approximately isotropic parameters, $t_x \approx t_y \equiv t$ and $V_x \approx V_y \equiv V$, thereby realizing the Hamiltonian in Eq.~\eqref{eq:hoppingHdelta}; the parameters used in the corresponding measurements are summarized in \tabref{tab:lattice_params}. To instead realize the Hamiltonian in Eq.~\eqref{eq:hoppingH}, we set the long-lattice depths to zero, such that only the short-lattice component of the potential is present.

\begin{table*}[!ht]
\centering
\caption{Summary of lattice parameters for measurements with the superlattice. The ratio $V/t$ is tuned by varying $V_{x,L}$ and $V_{y,L}$, while other parameters are kept approximately constant.
\label{tab:lattice_params}}
\begin{tabular}{@{}
  S[table-format=1.1(1)]
  S[table-format=1.1(1)]
  S[table-format=1.2(1)]
  S[table-format=1.2(1)]
  S[table-format=3(2)]
  S[table-format=3(2)]
  S[table-format=4(2)]
  S[table-format=4(2)]
  S[table-format=1.1]
@{}}
\toprule
{$V_{x,\mathrm{S}}$ (\si{\Erxs})} & {$V_{y,\mathrm{S}}$ (\si{\Erys})} & {$V_{x,\mathrm{L}}$ (\si{\Erxl})} & {$V_{y,\mathrm{L}}$ (\si{\Eryl})} & {$t_x/h$ (\si{\Hz})} & {$t_y/h$ (\si{\Hz})} & {$V_x/h$ (\si{\Hz})} & {$V_y/h$ (\si{\Hz})} & {$V/t$} \\
\midrule
6.4(3) & 6.2(3) & 0(0)    & 0(0)    & 309(25) & 308(24) & 0(0)    & 0(0)    & 0(0)   \\
6.4(3) & 6.2(3) & 0.13(1) & 0.13(1) & 309(25) & 308(24) & 191(10) & 181(9) & 0.6(0) \\
6.4(3) & 6.2(3) & 0.25(1) & 0.27(1) & 309(25) & 308(24) & 367(18) & 375(19) & 1.2(0) \\
6.4(3) & 6.2(3) & 0.38(2) & 0.4(2)  & 309(25) & 308(24) & 559(28) & 556(28)  & 1.8(0) \\
6.4(3) & 6.2(3) & 0.5(3)  & 0.53(3) & 309(25) & 308(24) & 735(37) & 736(37) & 2.4(0) \\
6.4(3) & 6.2(3) & 0.63(3) & 0.67(3) & 309(25) & 308(25) & 926(47) & 931(47) & 3(0)   \\
6.4(3) & 6.2(3) & 0.84(4) & 0.89(4) & 309(25) & 308(25) & 1235(62) & 1236(62) & 4(0)   \\
6.4(3) & 6.2(3) & 1.05(5) & 1.11(6) & 309(25) & 309(25) & 1542(78) & 1542(78) & 5(0)   \\
\bottomrule
\end{tabular}
\end{table*}

We calibrate the anti-symmetric phase using the following sequence: atoms are loaded into the lower-energy sites of a staggered superlattice potential along the direction to be calibrated, while the coupling along the orthogonal direction is turned off to create an array of isolated one-dimensional superlattices with frozen dynamics. We then modulate the long-lattice depth along the calibration direction and scan the modulation frequency to locate the resonance at which atoms are maximally transferred to the higher-energy sites, corresponding to the drive frequency matching the inter-site energy offset. The anti-symmetric phase is identified as the configuration that maximizes this resonance frequency~\cite{chenControllingCorrelatedTunneling2011}.

\subsection{Experimental sequences}
\label{sec:experimental_sequence}
All experimental sequences share the same overall structure. We prepare a spin-balanced, two-dimensional degenerate sample of $^{6}\mathrm{Li}$ atoms in the two lowest hyperfine states, $\lvert F = 1/2, m_F = \pm 1/2 \rangle$, confined to a single layer of a vertical optical lattice with a lattice constant of \SI{3}{\micro\m}. In-plane confinement is provided by a square box potential projected using a digital micromirror device, covering approximately $14 \times 14$ sites of the short lattice. The atoms undergo plain evaporation in the box trap and are subsequently loaded into the lattice following a protocol that depends on the specific experiment. The final atom number is controlled by adjusting the box-trap depth.

To realize an effective noninteracting Hamiltonian for spinless fermions, both spin states are retained during lattice loading at finite interaction strength to ensure thermalization. The interaction is then ramped to a weak value, approaching the noninteracting limit, before the state-preparation sequence concludes.

For detection, the two spin states are separated into adjacent layers of the vertical lattice~\cite{koepsell2020}. A single charge-pumping step increases the interlayer distance from $\SI{3}{\micro\m}$ to $\SI{9}{\micro\m}$. One layer is then removed to eliminate background fluorescence, and the remaining plane is imaged.

The lattice loading sequences for each measurement are described below.

\begin{figure}[!t]
\centering
\includegraphics[width=\columnwidth]{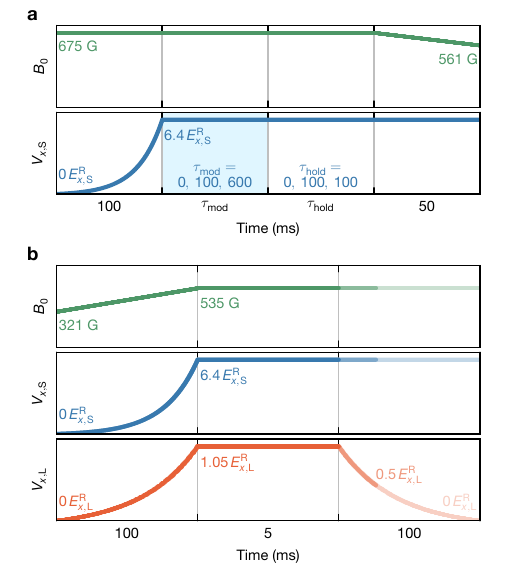}
\caption{
\textbf{Experimental lattice loading sequences.}
\textbf{(a)}~Loading sequence for the dephased Fermi liquid measurements. Elevated temperatures are generated via lattice modulation (shaded region). The modulation and hold durations for the three datasets at different temperatures are $(\tau_{\mathrm{mod}}, \tau_{\mathrm{hold}}) = (0,0)\,\mathrm{ms}$, $(100,100)\,\mathrm{ms}$, and $(600,100)\,\mathrm{ms}$.
\textbf{(b)}~Loading sequence for the SW-SSB phase transition measurements at $V/t = 5$, $2.4$, and $0$.
For clarity, only the $x$-lattices are shown; the corresponding $y$-lattice ramps closely overlap in both panels.
}
\label{fig:figS1-latticeloading}
\end{figure}

\paragraph{Dephased Fermi liquid.} For the experiments in Fig.\,\ref{fig:fig2-renyicorrelator}, atoms are loaded into a square lattice by simultaneously ramping the $x$- and $y$-short lattice depths to \SI{6.4}{\Erxs} ($t_x \approx \SI{309}{\Hz}$) and \SI{6.2}{\Erys} ($t_y \approx \SI{308}{\Hz}$) within \SI{100}{\milli\second} (Fig.\,\ref{fig:figS1-latticeloading}(a)). During the ramps, the magnetic field is set to $B_0 = \SI{675}{\Gauss}$, corresponding to an $s$-wave scattering length $\si{\as} = \SI{1155}{\ab}$ and yielding $U/t_x \approx U/t_y \approx 10$. After reaching the final lattice depths, the field is ramped to \SI{561}{\si{\Gauss}} in \SI{50}{\milli\second}, reducing the scattering length to $136.5\,\si{\ab}$. This sequence produces the dataset at $T = 0.45\,t$.

Datasets at $T = 0.61\,t$ and $T = 1.77\,t$ were generated via intraband heating by modulating the $x$- and $y$-short lattice depths at $\nu_{\textrm{mod}}=\SI{620}{\Hz} \approx 2t/h$ for \SI{100}{\milli\second} and \SI{600}{\milli\second}, respectively, after the lattices reach their final depths. The applied modulation corresponded to a relative modulation depth $V_{\mathrm{mod}}/V_{\mathrm{avg}}$ on the order of 10--20\%, ensuring efficient intraband heating. Once the modulation is turned off, the atoms are held in the lattice for an additional 100\,ms to thermalize before the interspecies scattering length is ramped down close to zero.

The vertical lattice with \SI{3}{\micro\m} spacing is kept at a depth of \SI{36}{\Erz} throughout the loading sequence, where \si{\Erz} denotes the associated recoil energy.

\paragraph{SW-SSB phase transition.} For the measurements in Fig.\,\ref{fig:fig3-swssb}, the lattice loading starts at a magnetic field $B_0 = \SI{321}{\Gauss}$, corresponding to a scattering length $\si{\as} = \SI{-290.3}{\ab}$. The $x$-short, $x$-long, $y$-short, and $y$-long lattices are simultaneously ramped within \SI{100}{\milli\second} to depths of \SI{6.4}{\Erxs}, \SI{1.05}{\Erxl}, \SI{6.2}{\Erys}, and \SI{1.11}{\Eryl}, respectively (Fig.\,\ref{fig:figS1-latticeloading}(b)). Concurrently, the magnetic field is increased to \SI{535}{\Gauss}, setting the scattering length to a small value of $\si{\as} = \SI{28.1}{\ab}$. These lattice depths correspond to $V/t = 5$.

For measurements performed at $V/t = 5$, the atoms are held in this configuration for \SI{5}{\milli\second}, after which the system is frozen for detection. For measurements at lower values of $V/t$, the depths of the long lattices are subsequently reduced to their final target values using an exponential ramp with a time constant of \SI{40.9}{\milli\second}.

Throughout the loading sequence, the vertical lattice with \SI{3}{\micro\m} spacing is maintained at a depth of \SI{45}{\Erz}.

In all sequences, a large initial scattering length is employed to ensure efficient plain evaporation in the box trap prior to lattice loading. For measurements of the SW-SSB phase transition, the system is initialized with attractive interactions, increasing the effective $V/t$ during lattice loading and thereby promoting double occupancy on the lower-energy sites.

\subsection{Data statistics}
\label{sec:data_statistics}
We acquired approximately 10,000 experimental snapshots at each temperature for the data shown in Fig.~\ref{fig:fig2-renyicorrelator}. Among these, more than 1,000 snapshots contain 27 particles in a $12 \times 12$ system, which are used for the plots in Fig.\,\ref{fig:fig2-renyicorrelator}(a). For the results shown in Fig.~\ref{fig:fig2-renyicorrelator}(b), where the size of the analysis region is varied, each data point is obtained from more than 1,000 samples with fixed system size and atom number.

For the measurements with varying $V/t$, we collected about 10,200 snapshots across eight different $V/t$ settings. From these data, we extract $6\times 6$ subregions with a fixed atom number of 9, yielding between 860 and 1,400 samples for each setting.

%%%%%%%%%%%%%%%%%%%%%%%%%%%%%%%%%%%%
% THEORETICAL DETAILS
%%%%%%%%%%%%%%%%%%%%%%%%%%%%%%%%%%%%

\section{Theory section}
\label{sec:suppmat:theory}
\subsection{Diagnosing SW-SSB}
${\rm U}(1)$ strong-to-weak spontaneous symmetry breaking (SW-SSB) in a quantum density matrix $\hrho$ is diagnosed by long-range order in the {\it fidelity correlator}:
\begin{equation}
    C^{(F)}(i,j)={\rm tr}\sqrt{\sqrt{\hrho}\hat{c}^\dagger_i\hat{c}_j\hrho \hat{c}^\dagger_j\hat{c}_i\sqrt{\hrho}}.
\end{equation}
In plain terms, this implies that moving a charge over large distances does not lead to a sharply distinguishable state. The fidelity correlator assesses this using the canonical measure of indistinguishability: the fidelity between $\hrho$ and $\hat{c}^\dagger_i\hat{c}_j\hrho \hat{c}^\dagger_j\hat{c}_i$. For the diagonal ensembles discussed in this work, the fidelity correlator is equivalent to the recently-proposed (and generally easier to measure) ``R\'{e}nyi-1" correlator:
\begin{equation}
    C^{(1)}(i,j)={\rm tr}\left(\sqrt{\hrho}\hat{c}^\dagger_i\hat{c}_j\sqrt{\hrho}\hat{c}^\dagger_j\hat{c}_i\right).
\end{equation}
Long-range order (LRO) in these correlation functions is ``stable" under finite-depth strongly-symmetric channels, e.g. any state exhibiting LRO in $C^{(F)}$ is not two-way connected by short-depth strongly-symmetric local channels to states without LRO. Thus,  these correlation functions serve as a diagnostic of mixed state phases.

Another diagnostic commonly used to probe SW-SSB is the ``R\'{e}nyi-2" correlator:
\begin{equation}
    C^{(2)}(i,j)=\frac{{\rm tr}\left(\hrho \hat{c}^\dagger_i\hat{c}_j\hrho \hat{c}^\dagger_j\hat{c}_i\right)}{{\rm tr}\hrho^2}.
\end{equation}
This correlation function directly considers two copies of the density matrix. While it does not satisfy the stability property of $C^{(F)}$ and $C^{(1)}$, it is often cited as a diagnostic because it is generally easier to calculate and serves as the simplest replica generalization of $C^{(1)}$. In the case of fully-dephased (diagonal) ensembles, moreover, it has been shown that $C^{(2)}(i,j)$ can be interpreted as the correlation function of a Gutzwiller-projected wavefunction in a doubled Hilbert space~\cite{su2025}.

\subsection{Quantum-Classical Correlators}
The fully-dephased density matrices we consider here may be written as
\begin{equation}
    \hrho=\mathcal{N}[\hrho_Q]=\sum_{\bf n}p_{\bf n}|\bf n\rangle\langle\bf n|
\end{equation}
where $\hrho_Q$ is the density matrix prepared in the quantum gas microscope, $\mathcal{N}$ is an infinite-strength density dephasing channel, and $\bf n$ are real-space configurations of fermions as measured experimentally with probabilities $p_{\bf n}=\langle{\bf n}|\hrho_Q|{\bf n}\rangle$. In terms of these, the R\'{e}nyi-$k$ correlators ($k=1,2$) are
\begin{eqnarray}
    C^{(k)}(i,j)&=&\frac{{\rm tr}\left((\mathcal{N}[\hrho_Q])^{k/2} \hat{c}^\dagger_i\hat{c}_j(\mathcal{N}[\hrho_Q])^{k/2} \hat{c}^\dagger_j\hat{c}_i\right)}{{\rm tr}\left(\mathcal{N}[\hrho_Q]\right)^k} \\ \label{eq:renyik}
    &=&\frac{\sum_{\bf n}p_{\bf n}^{k/2}p_{\bf n'}^{k/2}}{\sum_{\bf n}p_{\bf n}^{k}}
\end{eqnarray}
where the state $|{\bf n}'\rangle=\hat{c}^\dagger_i\hat{c}_j|\bf n\rangle$. These correlation functions can also be interpreted as standard two-point correlators with respect to positive semi-definite bosonic wavefunctions
\begin{equation}\label{eq:phi_renyik}
    |\phi^{(k)}\rangle=\frac{1}{\sqrt{\sum_{\bf n}p_{\bf n}^k}}\sum_{\bf n}p_{\bf n}^{k/2}|\bf n\rangle.
\end{equation}
This perspective provides an intuitive interpretation of SW-SSB in these systems, which is diagnosed as standard off-diagonal long-range order in the two-point function with respect to Eq.~\eqref{eq:phi_renyik}.

In general, quantum gas microscopes can only {\it sample} from $p_{\bf n}$ by performing ``snapshot" measurements of the density. This makes the measurement of these correlation functions a classically hard problem.
To circumvent this, we adopt the ``quantum-classical" correlation function~\cite{garratt2023} in which one of the density matrices is replaced by a classically-simulable state $\mathcal{N}[\hrho_C]$, which we will ``learn" from the experimentally-generated snapshots. %\cx{but with a new tweak thanks to machine-learning}. 
We refer to the classical configuration probabilities as $p^c_{\bf n}=\langle{\bf n}|\mathcal{N}[\hrho_C]|{\bf n}\rangle$. The quantum-classical (QC) correlators are then given by
\begin{eqnarray}\label{eq:qccorr}
    C^{(1)}_{\rm QC}(i, j)&=&\sum_{\bf n}p_{\bf n}\sqrt{\frac{p^c_{\bf n'}}{p^c_{\bf n}}} \\
    C^{(2)}_{\rm QC}(i, j)&=&\frac{\sum_{\bf n}p_{\bf n}p^c_{\bf n'}}{\sum_{\bf n}p_{\bf n}p^c_{\bf n}}.
\end{eqnarray}
In order to evaluate these expressions, we use the fact that $M$ experimentally-measured snapshots ${\bf n}_m$ with $m =1 \cdots M$ are sampled directly from the probability distribution $p_{\bf n}$. The QC correlators may thus be estimated:
\begin{eqnarray}\label{eq:qc1}
    C^{(1)}_{\rm QC}(i, j)&\approx&\frac{1}{M}\sum_m\sqrt{\frac{p^c_{{\bf n}'_m}}{p^c_{{\bf n}_m}}} \\ \label{eq:qc2}
    C^{(2)}_{\rm QC}(i, j)&\approx&\frac{\sum_m p^c_{{\bf n}'_m}}{\sum_m p^c_{{\bf n}_m}}.
\end{eqnarray}

For both correlation functions, we want to find a state $\hrho_C$ whose configuration probabilities closely approximate those of the quantum gas microscope. In the next section, we will discuss the general strategy we use to arrive at such approximations. In order to validate our approximation, we compare the behavior of the QC correlation function, $C^{(k)}_{\rm QC}$, to that of a ``classical-classical" (CC) correlation function, given by Eq.~\eqref{eq:renyik} with $p_{\bf n}\to p^c_{\bf n}$.
Provided that we can efficiently sample from $p^c_{\bf n}$, the CC correlator is classically simulable. We find that the condition ``QC=CC" is satisfied to within experimental error bars using the states defined below.

It will be important to work with strongly-symmetric density matrices $\mathcal{N}[\hrho_Q]$ and $\mathcal{N}[\hrho_C]$ in order to observe SW-SSB. The ensemble of snapshots ${\bf n}_m$ measured by the quantum gas microscope will not always have the same total particle number. As discussed in Sec.~\ref{sec:experiment}, particle number fluctuations due to the statistical loading process at finite temperature and technical noise are generally unavoidable.
All QC R\'{e}nyi correlators are thus produced using only a subset of the snapshots with a fixed total filling, i.e. those drawn from $\hat P_N\mathcal{N}[\hrho_Q](Z_{\rm GCE}/Z_N)$ where $\hat P_N$ is a projection onto the sector with $N$ total particles, $Z_{GCE}={\rm tr}(\hrho_Q)$, and $Z_{N}={\rm tr}(\hat P_N\hrho_Q)$. This projection into the canonical ensemble does not require any normalization of Eqs.~\eqref{eq:qc1} and~\eqref{eq:qc2} as sampling snapshots from $\hat P_N\mathcal{N}[\hrho_Q](Z_{\rm GCE}/Z_N)$ is equivalent to sampling from $\mathcal{N}[\hrho_Q]$ and postselecting on the total particle number.

Similarly, the classical states $\mathcal{N}[\hrho_C]$ described below will often be more natural to define in the grand canonical ensemble, whereby the canonical ensemble distribution is obtained by including a projection $\hat P_N$. In this case, one would find that $\sum_{\bf n}p^c_{\bf n}=1$ only if the sum includes configurations with differing particle numbers. If one restricts the sum to a particular charge sector, one must normalize the state accordingly so that the probabilities sum to 1. This normalization factor, however, trivially cancels from Eqs.~\eqref{eq:qc1} and~\eqref{eq:qc2}. Thus, it suffices to define $\mathcal{N}[\hrho_C]$ in the grand canonical ensemble and implement the strong-symmetry projection simply by postselecting the ensemble of snapshots.

\subsection{Learning Gaussian States}
\label{sec:gaussianlearning}
\subsubsection{General Remarks}
In this work we will take our classical states $\hrho_C$ to be Gaussian fermionic states. As the snapshots contain only the information of a single spin component within the quantum gas microscope, the state $\hrho_C$ will be a Gaussian state of spinless fermions. Such a state is generally defined as
\begin{equation}\label{eq:gce_rhoc}
    \hrho_C=\frac{1}{Z}\exp\left(-\sum_\alpha \omega_\alpha \hat{c}^\dagger_\alpha \hat{c}_\alpha \right)
\end{equation}
where $\hat{c}_\alpha$ is a fermionic annihilation operator in a single-particle eigenbasis state $\alpha$, $\omega_\alpha$ is the corresponding weight, and $Z$ is the normalization. The weights and basis vectors completely characterize the Gaussian state. Equivalently, the state is specified by the real-space correlation matrix, $G_{ij}={\rm tr}(\hrho \hat{c}^\dagger_i\hat{c}_j)$. The eigenvectors of $G$ constitute the single-particle eigenmodes (in real space), $\psi_\alpha(x)$,  while its eigenvalues may be interpreted as ``filling factors", $f_\alpha$, in the corresponding Fermi-Dirac distribution. These are related to the weights as $f_\alpha=1/(1+e^{\omega_\alpha})$.

Typically, a Gaussian fermionic state is defined with reference to a Boltzmann distribution, i.e. $\omega_\alpha=\beta(\epsilon_\alpha-\mu)$ where $\epsilon_\alpha$ is the energy of the single-particle state, $\beta=1/k_BT$ is the inverse temperature, and $\mu$ is a chemical potential. Indeed, if one knows $\mu$ and $\beta$, one can use the eigenvalues of $G$ to learn the single-particle energy spectrum; alternatively, given a single-particle spectrum, the eigenvalues of $G$ allow one to determine the temperature and chemical potential.

If the eigenvalues of $G$ are all $0$ or $1$, the state $\hrho_C$ corresponds to a Slater determinant and has a definite number of particles (given by ${\rm tr}~G$). As this represents a fermionic pure state, it can only ever be strongly-symmetric.
If $G$ has eigenvalues between $0$ and $1$, the Gaussian state is mixed. The density matrix $\hrho_C$ defined in Eq.~\eqref{eq:gce_rhoc} will then have an {\it indefinite} number of particles, corresponding to a {\it weakly symmetric} (grand-canonical) state. A canonical-ensemble state may generically be constructed by projecting a grand-canonical $\hrho_C$ into a state of definite particle number $N$ and normalizing the result.

\subsubsection{Learning}
In this work, we use two strategies for arriving at Gaussian states $\hrho_C$ that approximate $\hrho_Q$:
\begin{enumerate}
    \item We consider a thermal state whose eigenmodes $\psi_\alpha$ and eigenvalues $\epsilon_\alpha$ are fixed by an assumed form of the single-particle Hamiltonian. The temperature and chemical potential are {\it a priori} unknown.
    \item We consider Eq.~\eqref{eq:gce_rhoc} in full generality, without assuming the form of the eigenmodes or the values of the weights $\omega_\alpha$.
\end{enumerate}
We will focus here on case (2), which was used to compute all QC and CC correlators in the main text. We defer a discussion of case (1) to Sec.~\ref{sec:thermometry}.

We obtain the optimal $\hrho_C$ by minimizing the Kullback--Leibler (KL) divergence, $D_{\rm KL}$, between the quantum distribution $p_{\bf n}$ and the classical distribution $p^c_{\bf n}$:
\begin{equation}
    D_{\rm KL}(p||p^c)=\sum_{\bf n}p_{\bf n}\ln\frac{p_{\bf n}}{p^c_{\bf n}}.
\end{equation}
The KL divergence is a standard measure of distance between probability distributions. It is positive semi-definite and vanishes only when $p=p^c$. The utility of the KL divergence, relative to other measures of distance, is that one can minimize $D_{\rm KL}(p||p^c)$ with respect to $p^c$ given only a set of samples drawn from $p$. This can immediately be seen by expanding the logarithm and discarding the $p\ln p$ term, which is independent of $p^c$. The remaining term to be minimized is known as the {\it cross-entropy}, and its minimization is a standard task in machine learning applications. Given a collection of $M$ snapshots drawn from $p_{\bf n}$, we define a loss function
\begin{eqnarray}\label{eq:loss}
    L(\hrho_C)=-\frac{1}{M}\sum_m \ln p^c_{{\bf n}_m}.
\end{eqnarray}
By an extensive application of Wick's theorem, the configuration probabilities $p^c_{{\bf n}_m}$ may be expressed directly in terms of the real-space equal-time Green's function $G$~\cite{humeniuk2021}:
\begin{equation}\label{eq:canonical_prob}
    p^c_{\bf n}=(-1)^{N}\det\left(d_{\bf n}-G\right)
\end{equation}
where $d_{\bf n}$ is a diagonal matrix whose entries are $1-n_i$ on each site $i$ and $N=\sum_in_i$. In this way, the parameters of the Green's function may be optimized directly from the snapshot data.

Our unconstrained optimization begins with a general parameterization of an equal-time fermionic Green's function. The Green's function must (a) be Hermitian and (b) have eigenvalues $\lambda_i\in[0,1]$.
We therefore construct a simple generative model composed of a complex matrix $A$ and a real vector $x$. A unitary matrix $U$ is obtained by performing a complete, positive-definite QR decomposition of $A$, and a vector of Fermi-Dirac weights is constructed as $f_i=(1+\exp(x_i))^{-1}~\forall~x_i\in x$. The resulting Green's function is then given by $G=U^\dagger F U$ where $F$ is a diagonal matrix whose entries are the elements in $f$. 

Starting from a random initial state, the parameters of this model are iteratively updated to minimize the loss function in Eq.~\eqref{eq:loss}. We use stochastic gradient descent for efficiency and to escape local minima. We make use of native autodifferentiation tools in PyTorch in order to compute gradients of the loss function with respect to $A$ and $x$.

For training, we use the full distribution of snapshots produced by the quantum gas microscope, which spans different ${\rm U}(1)$ sectors. This produces a Green's function $G$ corresponding to a mixed state $\hrho_C$ with a weak ${\rm U}(1)$ symmetry. We restore the explicit strong ${\rm U}(1)$ symmetry by projecting into particular charge sectors. This proves to be necessary for two reasons: (1) We are able to train the model on an order of magnitude more snapshots ($\mathcal{O}(10,000)$) than we could if we post\-selected on the dominant charge sector, which produces more consistent results; and (2) The Gaussian distribution $p^c_{\bf n}$ that minimizes the KL divergence with respect to a strongly-symmetric dataset appears, generically, to correspond to that of a dephased {\it pure} state, $\mathcal{N}[|\psi\rangle\langle\psi|]$. Crucially, we find that these pure-state distributions violate the QC=CC condition, which is a necessary (although not sufficient) condition for approximating the true R\'{e}nyi correlators. The grand-canonical training, by contrast, produces probability distributions $p^c_{\bf n}$ that satisfy QC=CC as reported in the main text.

\subsection{Correlations on sub-systems}

In this work, the experimental ``region of interest" (ROI), i.e. the region in which the model Hamiltonians (Eqs.~\eqref{eq:hoppingH} and~\eqref{eq:hoppingHdelta}) are valid, is an approximately square region consisting of $14\times14$ lattice sites. All projective measurements from the quantum gas microscope are output as bit strings on this $14\times14$ grid.

Most of our results consider sub-regions within the ROI. This has a few advantages: (1) it reduces the (boundary) effects of the trapping and cooling scheme on our results; (2) it allows us to probe different system sizes with the same dataset; and (3) it increases the number of measurements at the desired filling fraction. This latter point is crucial: we must construct a strongly-symmetric (canonical) ensemble of snapshots in order to compute the QC R\'{e}nyi correlators. Our primary practical limitation in this work is the large sample complexity of these computations.

With this in mind, here we present the procedure we use for computing subsystem correlations. We will consider a generic dephased Gaussian state $\hrho=\sum_{\bf n}p^c_{\bf n}|\bf n\rangle\langle\bf n|$ defined on the full $14\times14$ ROI. As noted in Sec.~\ref{sec:gaussianlearning}, the probability of a configuration is given by Eq.~\eqref{eq:canonical_prob}.
Gaussian density matrices on a region $\mathcal{A}$ have the convenient property that the Green's function of a reduced density matrix on region $\mathcal{B}\in\mathcal{A}$ can be written as $G^{(B)}=G_{(i,j)\in B}$, i.e. the sub-matrix with indices $(i,j)$ inside $\mathcal{B}$. This Green's function $G^{(B)}$ defines all static properties measured in region $\mathcal{B}$ after tracing over the parts of $\mathcal{A} \notin \mathcal{B}$. Thus we may compute probabilities on a smaller $L_{\mathcal{B}}\times L_{\mathcal{B}}$ subregion of the ROI by simply constructing the Green's function on the subregion. This operation of tracing over the sites in $\mathcal{A}\notin\mathcal{B}$ generically results in a grand-canonical ensemble density matrix on $\mathcal{B}$, but as discussed earlier we may normalize the resulting probabilities such that they sum to unity within a given charge sector. Thus we may evaluate the {\it canonical} ensemble on generic $L_{\mathcal{B}}\times L_{\mathcal{B}}$ subregions of the ROI.

\subsection{Free fermion R\'{e}nyi correlators}

Here we discuss the calculation of free-fermionic R\'{e}nyi-$k$ correlators in Fig.~\ref{fig:fig2-renyicorrelator}(b). This calculation can be carried out exactly in the grand-canonical ensemble.

Let us consider a finite-temperature state $\hrho_\beta=e^{-\beta\hat H}/Z(\beta)$ where $Z(\beta)={\rm tr}(e^{-\beta\hat H})$. We will take the Hamiltonian $\hat H$ to be quadratic in the fermion creation and annihilation operators, and to be particle-number conserving. Our goal is to evaluate
\begin{equation}
    C^{(k)}_{ij}={\rm tr}\left(\hrho_\beta^{k/2}\hat c^\dagger_i\hat c_j\hrho_\beta^{k/2}\hat c^\dagger_j\hat c_i\right)/{\rm tr}\hrho_\beta^k
\end{equation}
We first note that $\hrho_\beta^{k/2}=e^{-\beta k\hat H/2}/Z^{k/2}(\beta)$. We then insert factors of $e^{-\beta k\hat H/2}e^{\beta k\hat H/2}$ in order to arrive at
\begin{equation}
    C^{(k)}_{ij}={\rm tr}\left(e^{-\beta k\hat H}\hat c^\dagger_i\hat c_j\hat c^\dagger_j(-\beta k/2)\hat c_i(-\beta k/2)\right)/Z(\beta k)
\end{equation}
where we define $\hat c_i(\tau)=e^{-\tau\hat H}\hat c_ie^{\tau\hat H}$. 

This now takes the form of an ordinary thermal expectation value at a temperature $\beta k$, so we can employ Wick's theorem. Making use of the notation $\langle\cdot\rangle_\beta={\rm tr}(e^{-\beta\hat H}\cdot)/Z(\beta)$, we find
\begin{multline}\label{eq:wickff}
    C^{(k)}_{ij}=\langle c^\dagger_ic_j\rangle_{\beta k}\langle c^\dagger_jc_i\rangle_{\beta k} \\ +\langle c^\dagger_ic_i(-\beta k/2)\rangle_{\beta k}\langle c_jc^\dagger_j(-\beta k/2)\rangle_{\beta k}
\end{multline}
Note that in the first term we made use of time-translational invariance. The first term is simply the modulus squared of the Green's function at temperature $\beta k$, which is short-ranged. The second term is the product of local Green's functions on sites $i$ and $j$, evaluated at different imaginary times. This term is independent of the separation between sites $i$ and $j$ and hence contributes to long-range order.

Equation~\eqref{eq:wickff} can be evaluated exactly in terms of the eigenvalues and eigenvectors of the Green's function. In Fig.~\ref{fig:fig2-renyicorrelator} we compute these from the spectrum of the machine-learned Green's functions, which we thus take to be an estimate of the R\'{e}nyi-$k$ correlators in the undephased experimental system.

\subsection{Relation between $\varrho^{(1)}$ and $\varrho^{(2)}$ in the localized limit}
Through the use of a superlattice potential, the quantum gas microscope can realize a highly localized state of $1/4$-filled spinless fermions. As noted in the main text, however, the experiment does not operate with a consistent thermal bath (i.e. a fixed temperature); rather, these localized states will have ``imperfections" that contribute to a finite entropy density, $S/N$. In this section we provide a schematic picture of the probability distribution of the experimental system at finite entropy density and in the localized limit, e.g. $V/t\gg 1$ in Eq.~\eqref{eq:hoppingHdelta}. This predicts a simple relation between the R\'{e}nyi-1 and R\'{e}nyi-2 condensate densities: $\varrho^{(2)}\propto\left(\varrho^{(1)}\right)^2$.

We will start with a simple ansatz for the probability distribution $p_{\bf n}$ in the highly localized regime. Here one expects that the configuration probabilities will be primarily local, e.g. $p_{\bf n}=\prod_jp_j(n_j)$ where $p_j(n_j)$ is the probability that site $j$ has density $n_j$. Noting that the fermion density is commensurate with the potential, we define $p_j=\delta_{1-n_j}p_0+\delta_{n_j}p_\delta$ if site $j$ is at a minimum of the superlattice potential and $p_j=\delta_{n_j}p_0+\delta_{1-n_j}p_\delta$ if it is not. We assume $p_0=(1-p_\delta)\gg p_\delta$. The interpretation of this ansatz is that each site has a most-likely density; the probability that the local density differs from this most-likely expectation is $p_\delta$. The probability of a configuration $\bf n$ is the product of these site-local probabilities.

This ansatz does not have a strong ${\rm U}(1)$ symmetry, but it nonetheless provides insight into the observed separation of scales for large $V/t$ in Fig.~\ref{fig:fig3-swssb}. The R\'{e}nyi-1 condensate density is then $\varrho^{(1)}=p_0p_\delta$, while the R\'{e}nyi-2 condensate density is $\varrho^{(2)}=p_0^2p^2_\delta/(p_0^2+p_\delta^2)^2$. For small $p_\delta$, one finds that $\varrho^{(2)}\propto\left(\varrho^{(1)}\right)^2$. Indeed we find that the ratio $\varrho^{(2)}_{\rm QC}/\left(\varrho^{(1)}_{\rm QC}\right)^2$ approaches an $\mathcal{O}(1)$ constant for large $V/t$, which is consistent with this prediction.

\section{Thermometry} \label{sec:thermometry}
In this section we show how we calculate the effective temperatures and entropy densities presented in the main text. In Sec.~\ref{sec:kl_thermometry} we compute temperatures by minimizing the KL divergence between the distribution of experimental snapshots and that of a thermal Gaussian state. This method was used to produce the results in Fig.~\ref{fig:fig3-swssb}(d). In Sec.~\ref{sec:g2thermometry} we present an alternative method based on fitting subsystem number fluctuations. We emphasize that these methods produce consistent temperature estimates for the quantum gas microscope.

\subsection{Minimizing KL divergence}
\label{sec:kl_thermometry}
Here we make use of the approach described in Sec.~\ref{sec:gaussianlearning} to compute the optimal thermal state approximation given a set of snapshots. Provided that we know the non-interacting Hamiltonian of the system, this serves as a {\it thermometer} for the quantum gas microscope.

We begin with an approximation of the single-particle Hamiltonian, $\hat{H}$. This is taken directly from the calibration of the quantum gas microscope, and generally takes the tight-binding form
\begin{equation}
    \hat{H}=-t\sum_{\langle ij\rangle}\hat{c}^\dagger_i\hat{c}_j+{\rm H.c.}+\sum_i\delta_i\hat{n}_i.
\end{equation}
We then diagonalize $\hat{H}$ to find its spectrum $\epsilon_\alpha$ and single-particle eigenstates $\psi_\alpha(x)$. A Gaussian Gibbs state with inverse temperature $\beta$ and chemical potential $\mu$ may then be constructed as
\begin{equation}
    \hrho_C(\beta,\mu)=\frac{1}{Z}\exp\left(-\beta\sum_\alpha(\epsilon_\alpha-\mu) \hat{c}^\dagger_\alpha \hat{c}_\alpha \right),
\end{equation}
where $\hat c_\alpha=\sum_x\psi_\alpha(x)\hat c_x$. In all of our thermal results, we either fix $\mu$ such that the average particle density of $\hrho_C$ matches that of the experimental system or we use the canonical ensemble (obviating $\mu$). Thus this is effectively a one-parameter model.

We seek to find the optimal inverse temperature $\beta$ that minimizes the KL divergence between $p_{\bf n}$, the experimental probability of measuring a configuration $\bf n$, and $p^c_{\bf n}(\beta)$, the associated classical (thermal) probability. This operation is equivalent to minimizing the cross entropy:
\begin{eqnarray}\label{eq:lossbeta}
    L(\beta)=-\frac{1}{M}\sum_m \ln p^c_{{\bf n}_m}(\beta).
\end{eqnarray}

As noted previously, the ensemble of experimentally-measured snapshots do not all have the same total charge. We may therefore perform this minimization in two ways: (a) in the canonical ensemble, by restricting the sum in Eq.~\eqref{eq:lossbeta} to snapshots with fixed particle number; or (b) in the grand canonical ensemble, leaving the sum unrestricted. Method (a) is more directly relevant to the measurement of SW-SSB, which as noted previously requires that the density matrices have an explicit strong symmetry. Method (a) also explicitly ignores total particle number fluctuations in the optical trap, which may not necessarily be ``thermal" (or at least may not be described by the same temperature as the state within the trap); it thus likely functions better as a ``thermometer" for non-interacting states in the quantum gas microscope than method (b). The advantage of method (b) is that one can use more snapshots to construct the loss function. While we focus on the results using method (a), it is perhaps interesting to note that these methods produce similar estimates for the optimal temperature.

Importantly, for canonical-ensemble optimization, one must normalize the probabilities by a constant $\beta$-dependent factor to ensure that the probabilities in a given charge sector are normalized: $\sum_{\bf n} p^c_{\bf n}=1$, with constraint $\sum_i n_i = N_{\rm tot}$. This is accomplished with the substitution $p^c_{\bf n}\to (Z_{\rm GCE}/Z_N)p^c_{\bf n}$ where $Z_{\rm GCE}$ is the grand-canonical partition function and $Z_N$ is the canonical partition function for an $N$-particle system.

The optimization is done manually by picking a grid of inverse temperatures $\{\beta_i\}$ and computing the value of Eq.~\eqref{eq:lossbeta} for each temperature. We then use a cubic spline interpolation to extract the optimal inverse temperature $\beta^*$. The uncertainty in the inverse temperature may be estimated from the cross entropy as
\begin{equation}
    \sigma_\beta\approx\left(M\frac{\partial^2L}{\partial\beta^2}\bigg|_{\beta=\beta^*}\right)^{-1/2}
\end{equation}

\subsection{Thermometry based on subsystem number fluctuations}
\label{sec:g2thermometry}

\paragraph{Fluctuation thermometry.}
For a noninteracting fermionic system in the grand-canonical ensemble, the occupations of different single-particle eigenmodes are independent and follow the Fermi--Dirac distribution $f_\alpha = [e^{(\epsilon_\alpha-\mu)/T}+1]^{-1}$. As a result, the total particle-number fluctuations are given by
\begin{equation}
\mathrm{Var}(\hat{N})
= \sum_{\alpha} \mathrm{Var}(\hat{n}_{\alpha})
= \sum_{\alpha} f_{\alpha}(1-f_{\alpha}),
\end{equation}
where $\alpha$ labels the single-particle eigenstates of the quadratic Hamiltonian and $\hat{n}_\alpha = \hat{c}^\dagger_\alpha \hat{c}_\alpha$. Since the occupations $f_\alpha$ depend on temperature, the particle-number fluctuations provide a direct probe of $T$ and are readily accessible in a quantum gas microscope (see also~\cite{dixmeriasFluctuationThermometryAtomresolved2025}).

The normalized fluctuations are given by the zero-momentum structure factor
\begin{equation}
S(0)
= \frac{\mathrm{Var}(\hat{N})}{N^{\textrm{site}}}
= \langle f_{\alpha}(1-f_{\alpha})\rangle_{\alpha},
\end{equation}
where $N^{\textrm{site}}$ is the number of lattice sites, and $\langle \cdot \rangle_{\alpha}$ denotes an average over single-particle eigenstates.

\begin{figure}[!t]
\centering
\includegraphics[width=\columnwidth]{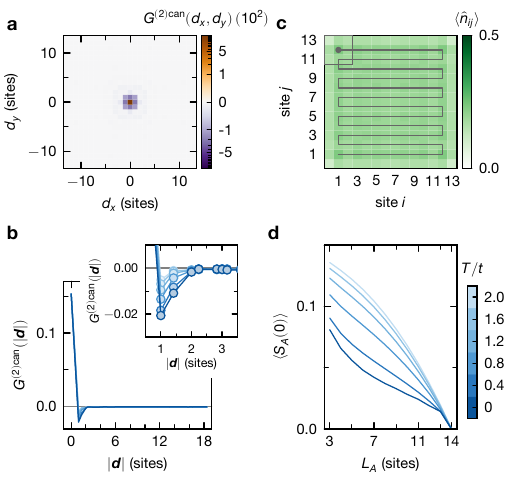}
\caption{
\textbf{Thermometry based on subsystem number fluctuations.}
All panels show theoretical results.
\textbf{(a)}~Connected density--density correlations 
$G^{(2)\mathrm{can}}(d_x,d_y)$ in real space for a canonical ensemble on a 
$14\times14$ square lattice with open boundary conditions at temperature 
$T=0\,t$ and filling $n \approx 0.189$. A pronounced Pauli hole at short 
distances is clearly visible.
\textbf{(b)}~Radially averaged correlations 
$G^{(2)\mathrm{can}}(|\boldsymbol{d}|)$ as a function of distance 
$|\boldsymbol{d}|$ for different temperatures $T/t$ spanning the range 
realized in the experiment. The inset highlights the short-distance behavior. 
The color encodes the temperature, with the corresponding color scale shown 
in \textbf{(d)}.
\textbf{(c)}~Local density profile $\langle \hat{n}_{ij} \rangle$ for the 
same system at temperature $T=0\,t$. The gray box indicates a subsystem of 
linear size $L_A=3$ sites used to compute $S_A(0)$. The gray line 
illustrates how this box is translated across the lattice in unit steps, 
yielding one value of $S_A(0)$ per position; these values are then 
averaged over all positions.
\textbf{(d)}~Position-averaged zero-momentum structure factor 
$\langle S_A(0) \rangle$ as a function of subsystem size $L_A$ for 
different temperatures $T/t$. The shown range spans the temperatures realized 
in the experiment and provides the basis for thermometry.
}
\label{fig:temperaturecalibrationcurves}
\end{figure}

\paragraph{Canonical ensemble and subsystem fluctuations.}
While the total particle-number fluctuations $\mathrm{Var}(\hat{N})$ provide a natural thermometric probe in the grand-canonical ensemble, in a quantum gas microscope shot-to-shot fluctuations can partly arise from technical noise, making $\mathrm{Var}(\hat{N})$ suboptimal for thermometry. We therefore fix the total atom number $N$ by postselecting experimental snapshots, thereby imposing a canonical constraint, and instead consider atom-number fluctuations within subsystems.

The fluctuations within a subsystem are determined by real-space density--density correlations, described by the connected correlator in the canonical ensemble,
\begin{equation}
G^{(2)\mathrm{can}}_{ij}
=
\langle \hat{n}_i \hat{n}_j \rangle_N
-
\langle \hat{n}_i \rangle_N \langle \hat{n}_j \rangle_N.
\end{equation}
Specifically, for a subsystem $A$ of linear size $L_A$ (with $N^{\textrm{site}}_A = L_A^2$ sites), the particle-number fluctuations can be written as
\begin{equation}
S_A(0)
=
\frac{1}{N^{\textrm{site}}_A}\sum_{i,j\in A} G^{(2)\mathrm{can}}_{ij}
=
\frac{\mathrm{Var}(\hat{N}_A)}{N^{\textrm{site}}_A},
\end{equation}
where $\hat{N}_A=\sum_{i\in A}\hat{n}_i$. As the subsystem size is varied, the sum over $i,j\in A$ weights correlations at different distances differently, leading to a size dependence of $S_A(0)$ that reflects the spatial structure of $G^{(2)\mathrm{can}}_{ij}$. This size dependence provides a sensitive probe of temperature and forms the basis of our thermometry scheme. In particular, while the real-space correlator $G^{(2)\mathrm{can}}(|\boldsymbol{d}|)$ is nonzero only at short distances and exhibits small absolute variations with temperature, the integrated quantity $S_A(0)$ accumulates contributions over many site pairs. As a result, its dependence on temperature is significantly enhanced, leading to improved sensitivity (see Fig.~\ref{fig:temperaturecalibrationcurves}).

In practice, for each subsystem size, we slide a region $A$ across the system with unit stride in both spatial directions and average over all positions, obtaining the spatially averaged quantity $\langle S_A(0)\rangle$, where $\langle \cdot \rangle$ denotes an average over subsystem positions.

In theory, $\langle S_A(0)\rangle$ is obtained by computing $S_A(0)$ from $G^{(2)\mathrm{can}}_{ij}$ and averaging over all subsystem positions, while in experiment it is obtained from the fluctuations $\mathrm{Var}(\hat{N}_A)$ measured within the same regions and averaged in the same way. The experimentally obtained values, based on snapshots postselected to fixed atom number $N$, are then compared to theoretical curves of $\langle S_A(\mathbf{0})\rangle$ as a function of subsystem linear size $L_A$ to determine the temperature.

\paragraph{Model implementations.}

\begin{figure}[!t]
\centering
\includegraphics[width=\columnwidth]{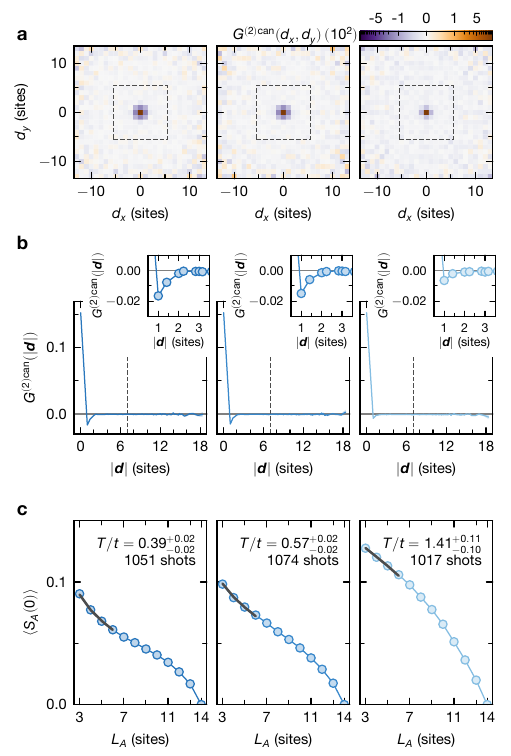}
\caption{
\textbf{Temperature extraction from $S_A(0)$ in the homogeneous system.}
\textbf{(a)}~Connected density--density correlations 
$G^{(2)\mathrm{can}}(d_x,d_y)$ in real space for three experimental datasets 
on a $14\times14$ square lattice, obtained from snapshots postselected to a 
fixed total atom number $N=37$. From left to right, the panels correspond to datasets at increasing temperatures. The gray dashed square indicates the spatial range 
of correlations included in the thermometry fits.
\textbf{(b)}~Radially averaged correlations 
$G^{(2)\mathrm{can}}(|\boldsymbol{d}|)$ corresponding to (a). The dashed 
vertical line indicates the cutoff used for fitting. 
Lines connect the experimental data points as guides to the eye, while the 
shaded region indicates the statistical uncertainty. The inset highlights the short-distance behavior, and error bars are smaller than the markers where not visible.
\textbf{(c)}~$S_A(0)$ as a function of subsystem size $L_A$. Error analysis is performed via bootstrapping over experimental snapshots, and error bars and uncertainties denote 95\% confidence intervals. Symbols denote experimental data, with error bars smaller than markers where not visible. Solid blue lines show the median. Solid gray lines show the theoretical curves corresponding to the median fitted temperatures.
}
\label{fig:temperaturefitflatsystem}
\end{figure}

\begin{figure}[!t]
\centering
\includegraphics[width=\columnwidth]{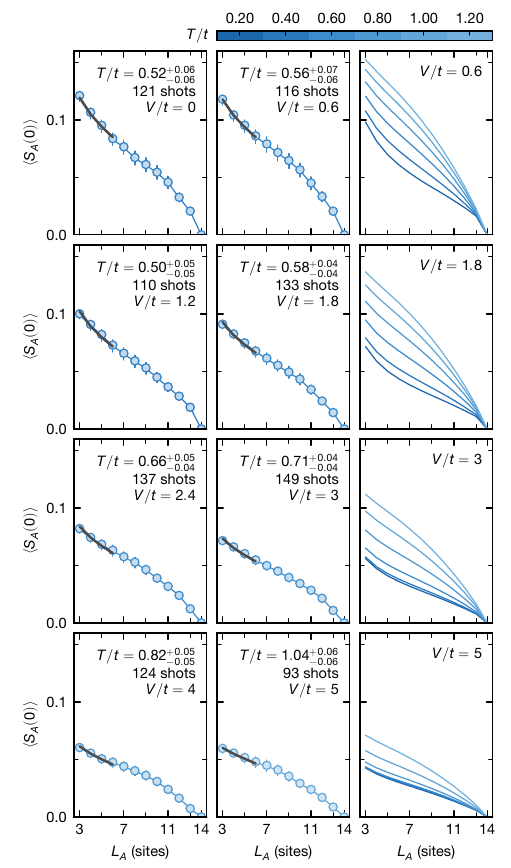}
\caption{
\textbf{Temperature extraction from $S_A(0)$ for varying $V/t$.}
$S_A(0)$ as a function of subsystem size $L_A$ for experimental 
datasets at different values of $V/t$ (two leftmost columns). Error analysis is performed via bootstrapping over experimental snapshots, and error bars and uncertainties denote 95\% confidence intervals. Symbols denote experimental data, with error bars smaller than markers where not visible. Solid blue lines show the median. Solid gray lines show the theoretical curves corresponding to the median fitted temperatures. The rightmost column shows theoretical 
results for $S_A(0)$ at four values of $V/t$, computed on a 
$14\times14$ lattice with open boundary conditions at filling 
$n \approx 0.28$. At large $V/t$ and low $T/t$, the temperature dependence of $S_A(0)$ is weaker.
}
\label{fig:temperaturefitVt}
\end{figure}

We now apply the above formalism to two Hamiltonians realized in the experiment.

For the measurements presented in the ``Dephased Fermi liquid'' section, we analyze experimental snapshots postselected to contain $N=37$ atoms within a $14\times14$ region, for low-, intermediate-, and high-temperature datasets, with the temperature determined independently in each case. The theoretical curves are computed for the same system size and particle number, using a square lattice with open boundary conditions described by the Hamiltonian in Eq.~\eqref{eq:hoppingH}.

To determine the temperature, we compare the measured $\langle S_A(0) \rangle$ as a function of subsystem linear size $L_A$ to the theoretical curves (see Fig.~\ref{fig:temperaturefitflatsystem}). Rather than using the full range of subsystem sizes, we restrict the fits to $L_A \leq 6$, which corresponds to including correlations up to a maximal distance $|\boldsymbol{d}| \leq 5\sqrt{2}$.

This restriction is motivated by boundary effects. In the theoretical model with open boundary conditions, a density depletion occurs at the edges (see Fig.~\ref{fig:temperaturecalibrationcurves}), whereas in the experiment the boundary is not described by this idealized condition due to the finite steepness of the optical box trap, and the system size is defined by sites with significant density. As $L_A$ increases, boundary contributions become more prominent, leading to deviations between theory and experiment. Restricting to smaller $L_A$ therefore ensures a reliable comparison.

The extracted temperatures are slightly lower than those reported in the main text, which were obtained using the method presented in Sec.~\ref{sec:kl_thermometry}, but the discrepancy remains modest.

For the measurements shown in the ``SW-SSB phase transition'' section, we analyze experimental snapshots within a $14\times14$ system for different values of $V/t$, with the temperature extracted independently for each. The data are postselected to have a fixed total atom number $N=55$, giving a filling of $n\approx0.28$. This choice yields approximately 100 experimental shots per value of $V/t$, resulting in sufficiently low-noise $S_A(0)$ curves. The corresponding theoretical curves are computed for the same system size and particle number, using a square lattice with site-dependent energy offsets described by the Hamiltonian in Eq.~\eqref{eq:hoppingHdelta}. As in the homogeneous case, the fits are restricted to $L_A \leq 6$. The extracted temperatures are in good overall agreement with those reported in the main text, which were obtained using the method described in Sec.~\ref{sec:kl_thermometry}. We note that the temperatures reported in Fig.~\ref{fig:fig3-swssb} were extracted at a slightly different filling of $n=0.25$.

As shown in Fig.\,\ref{fig:fig3-swssb}, increasing $V/t$ drives the system into a regime with strongly suppressed low-energy density fluctuations. In this regime, at low temperatures, the density distribution becomes effectively pinned, and $\langle S_A(0)\rangle$ loses its sensitivity to temperature, as illustrated in Fig.\,\ref{fig:temperaturefitVt}. This limits the applicability of the fluctuation-based thermometry in this regime.

\section{Overview of SW-SSB} \label{sec:overview}

\begin{table*}[!ht]
\centering

\setlength{\tabcolsep}{.1cm}
\renewcommand{\arraystretch}{2.5}
\begin{tabular}{%
    c c c c c c
}
\toprule
{} 
& 
{$\mathrm{tr}(\hrho \hat{b})$}
&
{$\displaystyle\lim_{x\to\infty} \mathrm{tr}(\hrho \hat{b}_0 \hat{b}_x^{\dagger})$}
&
\makecell{$\dfrac{\mathrm{tr}(\hrho^{k/2} \hat{b}^\dagger \hrho^{k/2} \hat{b})}{\mathrm{tr}(\hrho^k)}$
\\[2mm]
\graynote{$\displaystyle
\frac{\langle\!\langle \hrho^{k/2} \vert \Delta \vert \hrho^{k/2} \rangle\!\rangle}
{\langle\!\langle \hrho^{k/2} \vert \hrho^{k/2} \rangle\!\rangle}$}
}
&
\makecell{$\displaystyle\lim_{x\to\infty} \dfrac{\mathrm{tr}(\hrho^{k/2} \hat{b}_0 \hat{b}_x^\dagger \hrho^{k/2} \hat{b}_0^\dagger \hat{b}_x)}{\mathrm{tr}(\hrho^k)}$
\\[2mm]
\graynote{$\displaystyle
\lim_{x\to\infty}
\frac{\langle\!\langle \hrho^{k/2} \vert \Delta_0 \Delta_x^\dagger \vert \hrho^{k/2} \rangle\!\rangle}
{\langle\!\langle \hrho^{k/2} \vert \hrho^{k/2} \rangle\!\rangle}$}
}
&
{Examples}
\\
\midrule
{Strong symmetry} & {$0$} & {$0$} & {$0$} & {$0$} & \makecell{Pure quantum system\\(fixed particle number)} \\
% \midrule
{Weak symmetry} & {$0$} & {$0$} & {$\neq 0$} & {$\neq 0$} & \makecell{Classical GCE
%ensemble\\(grand canonical)
} \\
% \midrule
{No symmetry} & {$\neq 0$} & {$\neq 0$} & {$\neq 0$} & {$\neq 0$} & \makecell{System without\\particle number conservation} \\
% \midrule
{ST-SSB} & {$0$} & {$\neq 0$} & {$0$} & {$\neq 0$} & \makecell{Superfluid (CE)} \\
% \midrule
{WT-SSB} & {$0$} & {$\neq 0$} & {$\neq 0$} & {$\neq 0$} & \makecell{Superfluid (GCE)} 
%\\(grand canonical ensemble)} 
\\
% \midrule
{SW-SSB} & {$0$} & {$0$} & {$0$} & {$\neq 0$} & \makecell{Fully dephased\\Fermi liquid state} 
\\[2mm]

\bottomrule
\end{tabular}
\caption{Summary of six scenarios of symmetry and symmetry breaking, including their diagnostics: 1. systems with strong symmetry, 2. systems with weak symmetry, 3. systems with no symmetry or explicitly symmetry breaking (ESB), 4. ordinary spontaneous symmetry breaking (SSB) in quantum systems, now called ``strong-trivial spontaneous symmetry breaking" (ST-SSB), 5. ordinary spontaneous symmetry breaking in classical grand canonical ensemble (GCE), now called ``weak-to-trivial spontaneous symmetry breaking" (WT-SSB), 6. the strong-to-weak spontaneous symmetry breaking (SW-SSB) discussed in this work.}
\label{tab:sixscenarios}
\end{table*}

\subsection{Standard Symmetry and Symmetry breaking in pure quantum states}

In this section we provide a self-contained pedagogical overview of strong-to-weak spontaneous symmetry breaking (SW-SSB). Let us first review the standard notion of symmetry and spontaneous symmetry breaking (SSB) in pure quantum states. We will always use $\U(1)$ symmetry as our example, and consider bosonic systems first.

We start with a $0d$ quantum system, which is just a single site. The $\U(1)$ symmetry action is a unitary operator \beqn \hat G = e^{\ii \theta \hat{n}}, \eeqn where $\hat{n}$ is the particle number operator. We say a quantum state $|\Psi\rangle$ on this site is symmetric under $\U(1)$ when the state is invariant under the operation $\hat G$. Note that in quantum mechanics, a state is considered invariant under a transformation if it changes up to a global phase factor. For example, any eigenstate of $\hat{n}$, i.e. $\hat n| n \rangle=n|n\rangle $, is invariant under $\hat G$, as \beqn \hat G| n \rangle = e^{i \theta n } | n \rangle. \eeqn

One can also construct a state that is not symmetric under the U(1) symmetry, meaning under $\hat G$ it changes by more than a pure phase factor. For example, let us consider a state \beqn | + \rangle = \alpha |0\rangle + \beta |1\rangle. \eeqn Under $\hat G$, the state $|+\rangle$ transforms as \beqn \hat G|+\rangle = \alpha |0\rangle + e^{\ii \theta} \beta |1\rangle \neq e^{\ii \theta} |+\rangle. \eeqn Therefore, $|+\rangle$ is not symmetric under $\U(1)$ (or equivalently, $|+\rangle$ does not have the $\U(1)$ symmetry).

The fact that $|+\rangle$ breaks the $\U(1)$ symmetry can be diagnosed through the following expectation value \beqn \langle + | \, \hat{b} \, | + \rangle \neq 0. \eeqn In fact, we can use this as the definition for {\it explicit} $\U(1)$ symmetry breaking: explicit $\U(1)$ symmetry breaking means the expectation value of the boson creation/annihilation operator is nonzero.

In a quantum many-body system, like the Bose-Hubbard model, there is a superfluid phase where the $\U(1)$ symmetry is spontaneously rather than explicitly broken. For a large system, the $\U(1)$-SSB is diagnosed by the following long-range correlation \beqn && \lim_{x \ra \infty} C(x) \neq 0, \cr \cr && C(x) = \langle \Psi | \, \hat{b}_{0} \, \hat{b}_x^\dagger \, | \Psi\rangle = \tr\left(\, \hrho \, \hat{b}_{0} \hat{b}_x^\dagger \, \right). \eeqn In the first-quantized language, $C(x)$ is the off-diagonal correlation function: \beqn C(x) = \int \prod_{i = 2}^N d^d x_i \Psi^\ast(0, x_2 \cdots x_N) \Psi(x, x_2, \cdots x_N ). \eeqn When $C(x)$ saturates to a constant with large $x$, we say the system has off-diagonal long range order (ODLRO); if $C(x)$ only decays algebraically rather than exponentially with large $x$, the system has off-diagonal {\it quasi} long range order. In both cases we say the U(1) symmetry is spontaneously broken. 

Note that the particle number is fixed in a superfluid, therefore $\langle \Psi | \, \hat{b} \, | \Psi\rangle$ must vanish; but it still behaves like a system with symmetry breaking because of nonzero $\lim_{x \ra \infty} C(x)$. It is in this sense that the U(1) symmetry is broken, but only {\it spontaneously} rather than {\it explicitly} broken.

\subsection{Symmetry of Density matrices}

When discussing density matrices, the notion of ``symmetry" is further expanded. Let us still consider a single site, and compare the following three density matrices: 
\beqn 
\hrho_s &=& |0\rangle \langle 0|, \cr \cr 
\hrho_+ &=& |+\rangle \langle + |, \ \ \ \ {\rm with} \ \ \ \ |+\rangle = \alpha |0\rangle + \beta |1\rangle \cr\cr
\hrho_w &=& p_0 |0\rangle \langle 0| + p_1 |1\rangle \langle 1 |, \ \ \ {\rm and} \ \ p_0 + p_1 = 1. \label{3ex} 
\eeqn 
These three density matrices each have their own characteristics under $\hat G$:

\begin{itemize}

\item $\hrho_s$ has the {\bf strongest symmetry} (or strongest invariance): 
\beqn 
\hat G \hrho_s = \hrho_s, \ \ \ \ {\rm and} \ \ \ \hrho_s \hat G^\dagger = \hrho_s. 
\eeqn 
This means that $\hrho_s$ is invariant under left and right actions of $\hat G$ separately. Because $\hat G$ can act separately from left and right side of the density matrix, the strong symmetry can also be called a ``doubled symmetry". This ``doubled" symmetry becomes more explicit in the Choi doubled-space representation of the density matrix, which we will discuss later.

\item $\hrho_w$ has a {\bf ``weaker" symmetry}: 
\beqn 
&& \hat G \hrho_w \neq \hrho_w, \ \ \ \hrho_w \hat G^\dagger \neq \hrho_w, \cr\cr 
{\rm but} && \hat G \hrho_w \hat G^\dagger = \hrho_w.  
\eeqn 
Because $\hrho_w$ is only symmetric under simultaneous left and right action of $\hat G$, the weak symmetry is also sometimes called ``diagonal symmetry".

\item $\hrho_+$ simply has no symmetry at all, as $ \hat{G} \hrho_+ \hat G^\dagger \neq \hrho_+$.

\end{itemize}

Let us make some general remarks of these symmetries:

\begin{itemize}

\item {\bf (1)} The density matrix $\hrho_w$ has ``weak symmetry", and it is a mixed rather than pure density matrix. In fact, the notion of weak symmetry, as well as strong-to-weak spontaneous symmetry breaking (SW-SSB) that we will discuss soon, often arises when a quantum system loses it quantum coherence. Therefore SW-SSB is a feature often associated with the process of decoherence. This is also consistent with our observation that increasing temperature would enhance rather than suppress the SW-SSB (The strength of the inter-replica Cooper pair correlation in the doubled space).

\item {\bf (2)} We can also examine the origin of different symmetries of a density matrix:

{\bf (2.1)} Given a density matrix $\hrho$, if every eigenstate of $\hrho$ is also an eigenstate of total particle number $\hat{N}$, and all eigenstates have the {\it same eigenvalue} of $\hat{N}$, then the density matrix has the strong U(1) symmetry.

Physical example: if a thermal density matrix is in the canonical ensemble (CE), i.e. $\hrho \sim e^{- \beta \hat{H}}$ with fixed total particle number, it has the strong symmetry, as every eigenstate of the density matrix is the eigenstate of the total particle number, and all the eigenstates have the same total particle number.

{\bf (2.2)} If every eigenstate of a density matrix is an eigenstate of total particle number $\hat{N}$, but different eigenstates have {\it different eigenvalues} of $\hat{N}$, then the density matrix has the weak U(1) symmetry.

Physical example: if a thermal density matrix is in the grand canonical ensemble (GCE), i.e. $\hrho \sim e^{- \beta (\hat{H} - \mu \hat{N})}$, it has the weak symmetry, as every eigenstate of the density matrix is the eigenstate of the total particle number, but different eigenstates can have different total particle number $\hat{N}$.

{\bf (2.3)} If the eigenstate of a density matrix is {\it NOT} the eigenstate of total particle number $\hat{N}$, then the density matrix does not have any symmetry.

\item {\bf (3)} A pure density matrix with fixed particle number $\hat{N} = N$ (i.e. it has the strong $\U(1)$ symmetry) can become a mixed state through interacting with the environment (ancilla degrees of freedom), followed by ``tracing out" the environment. 

Let's assume the system interacts with the environment with the following Hamiltonian: \beqn \hat{H}_{int} = \lambda \hat{O}_s \hat{O}_a + h.c. \eeqn
If $\hO_s$ and $\hO_a$ are both invariant under $\U(1)$, e.g. they are (for example) energy density operators, then the reduced mixed density matrix of the system after tracing out the environment still has the strong U(1) symmetry. In this case, the particle number of the system and environment are separately conserved.

If $\hO_s$ and $\hO_a$ are not invariant under $\U(1)$, but the entire $\hat{H}_{int}$ is invariant under $\U(1)$, then the resulting reduced mixed density matrix of the system will have the weak U(1) symmetry. For example, we can take $\hO_s = \hat{b}^\dagger_s$, $\hO_a = \hat{b}_a$. In this case, the system and
environment exchanges particle numbers, while preserving the total
particle number of the system and the environment.

\end{itemize}

The symmetries of the density matrices can still be diagnosed through expectation values, but in order to clearly distinguish symmetries of density matrices, we will need to use ``R\'{e}nyi" expectation values, i.e. quantities nonlinear with density matrices. For the three simple example density matrices in Eq.~\eqref{3ex}, one can easily verify that 
\beqn 
&& \tr\left( \hrho_+ \hat{b} \right) \neq 0, \ \ \ \ \tr\left( \hrho_s \hat{b}
\right) = \tr\left( \hrho_w \hat{b} \right) = 0. \cr\cr 
&& \frac{\tr\left( \hrho_s \hat{b}^\dagger \hrho_s \hat{b} \right)}{\tr \hrho_s^2} =
0, \ \ \ \frac{\tr\left(\hrho_w \hat{b}^\dagger \hrho_w \hat{b} \right)}{\tr
\hrho_w^2} \neq 0. 
\label{swdef} 
\eeqn

From these equations, we can see that quantities linear in the density matrix cannot distinguish density matrices with strong and weak symmetry (because $\tr\left( \hrho_s \hat{b} \right)$ and $ \tr\left( \hrho_w \hat{b} \right)$ both vanish); one needs observables nonlinear in the density matrix to diagnose the strong or weak symmetry.

\subsection{Strong-to-Weak SSB}

The weak symmetry is a subgroup of strong symmetry, therefore it is possible for a system with strong symmetry to be spontaneously broken to weak symmetry.

As we explained before, in the standard story of symmetry and symmetry breaking, if a quantum state $|\Psi\rangle$ has a U(1) symmetry, but there is a U(1)-SSB, it means \beqn && \langle \Psi | \, \hat{b} \, |\Psi\rangle = 0, \cr\cr && \lim_{x \ra \infty} C(x) = \lim_{x \ra \infty} \langle \Psi | \, \hat{b}_0 \, \hat{b}_x^\dagger \, |\Psi\rangle \neq 0. \eeqn Since we have expanded the notion of symmetry to strong and weak, the SSB described above for a pure quantum system is a full breaking of the strong U(1) symmetry, or breaking from the strong U(1) symmetry to its trivial subgroup, i.e. strong-to-trivial spontaneous symmetry breaking (ST-SSB). 

A direct analogy is that, if a density matrix $\hrho$ has strong U(1) symmetry, but there is a SW-SSB, it means \beqn && \frac{\tr\left( \hrho \, \hat{b}^\dagger \, \hrho \, \hat{b} \right)}{\tr \hrho^2} = 0, \ \ {\rm but}  \ \ \lim_{x \ra \infty} C^{(2)}(x) \neq 0, \cr \cr && C^{(2)}(x) = \frac{\tr\left( \hrho \, \hat{b}_0 \hat{b}^\dagger_x \, \hrho \,
\hat{b}_0^\dagger \hat{b}_x \right)}{\tr \hrho^2}. \eeqn
The last quantity is called the R\'{e}nyi-2 correlator. More generally one can define a series of R\'{e}nyi-$k$ correlators nonlinear with the density matrix: \beqn  C^{(k)}(x) = \frac{\tr\left( \hrho^{k/2} \, \hat{b}_0 \, \hat{b}^\dagger_x \, \hrho^{k/2} \, \hat{b}_0^\dagger \, \hat{b}_x \right)}{\tr \hrho^k}. \eeqn 
The most precise diagnostic of SW-SSB is the R\'{e}nyi-1 correlator (or equivalently the fidelity correlator). The R\'{e}nyi-2 correlator is the simplest replica generalization one often adopts in theoretical studies. In our experiment, we measure {\it both} R\'{e}nyi-$1,2$ correlators.

Let us consider a density matrix with strong symmetry $\hat G$, meaning it satisfies $\hat G\hrho \sim \hrho \hat G^\dagger \sim \hrho$ (here $\sim$ means identical up to a phase factor). This strong symmetry $\hat G$ can either be partially broken down to weak symmetry (SW-SSB) or completely broken to trivial symmetry (ST-SSB). These two scenarios are characterized by two different correlation functions: 
\beqn 
{\rm SW-SSB} &:& \lim_{x \ra \infty}
%\frac{\tr\left( \hrho \ b_0 b^\dagger_x \hrho b_0^\dagger b_x
%\right)}{\tr \hrho^2}
C^{(1)}(x) \neq 0, \cr\cr {\rm ST-SSB} &:& \lim_{x \ra \infty}
%\tr\left( \hrho \ b_0 \ b^\dagger_x \right)
C(x) \neq 0.  
\eeqn 
In this language, the standard SSB such as the SSB in a superfluid with a fixed particle number, is the ST-SSB scenario.

There is a class of density matrices of particular relevance to our problem. Let's consider a density matrix $\hrho$ completely {\it diagonal} in the density configurational space: 
\beqn 
\hrho = \sum_{\bf{n}} p_{\bf{n}} |{\bf{n}} \rangle \langle {\bf{n}}|. 
\eeqn 
For these diagonal density matrices, we can view $p^{k/2}_{\bf{n}}$ as a wave function $\Phi_k$, then the R\'{e}nyi correlators become 
\beqn 
&& C^{(k)}(x) = \cr\cr && \int \prod_{i = 2}^N d^dx_i \Phi_k^\ast(0, x_2, \cdots x_N) \Phi_k(x, x_2, \cdots x_N). 
\eeqn

With these preparations, we can consider a few example systems:

\begin{itemize}

\item If $p_{\bf{n}}$ is the probability distribution of a crystalline insulator wave function, i.e. $\hrho$ is the density matrix of a fully dephased trivial insulator, then $C^{(k)}(x)$ is short-ranged, i.e \beqn C^{(k)}(x) \sim \exp(- |x|/\xi_k). \eeqn The system is fully strongly symmetric, i.e. there is no SW-SSB.

\item If $p_{\bf{n}}$ is the probability distribution of a metal (a Fermi surface state) in dimensions higher than one, then $C^{(k)}(x)$ is long-ranged, i.e a fully dephased metal has the SW-SSB. We have shown both examples above in our experiment.

\item {\bf A bit of surprising example}: if $p_{\bf{n}}$ is the probability distribution of a $2d$ Chern insulator, or fractional quantum Hall (FQH) state, then $C^{(k)}(x)$ also decays with a power-law, i.e the fully dephased $2d$ quantum Hall state has a quasi-long range SW-SSB. A quantum Hall state (integer or fractional) is described by the Laughlin wave function \beqn \Psi(z_i) \sim \prod_{i<j} (z_i - z_j)^m. \eeqn Our ``probability wave function" $\Phi_k$ for the Laughlin state is \beqn \Phi_k(z_i) \sim \prod_{i < j} |z_i - z_j|^{m k}. \eeqn This is precisely the ``composite boson" wave function of the FQH state, and it is known that this wave function has quasi long range order~\cite{girvin1987}. This means that a fully dephased quantum Hall state should have quasi-long range SW-SSB, as has been predicted recently~\cite{sarma2025,wang2025a}.

\end{itemize}

\subsection{The ``Choi" (doubled space) representation}
\label{sec:supmat:choi}

The ``Choi" (doubled space) representation makes the symmetry and symmetry breaking of a density matrix very explicit. Let us start with a pure quantum state with a U(1) symmetry, its density matrix $\hrho_0 = |\Psi_0\rangle \langle \Psi_0|$ has the strong U(1) symmetry. To make a direct connection with our experiment, we assume $|\Psi_0\rangle$ is a spinless fermion state. In its doubled space representation, the density matrix $\hrho_0$ reads 
\beqn 
|\hrho_0\rrangle = |\Psi_0\rangle_L \otimes |\Psi^\ast_0\rangle_R. 
\eeqn 
The left and right replica space of the Choi representation corresponds to the ket and bra states of the original density matrix. If the original density matrix is a pure spinless Fermi surface state, the left and right replica spaces are decoupled, and the doubled state is effectively a spin-1/2 Fermi surface state.

The strong symmetry becomes two separate U(1) symmetries acting on the left and right replica spaces: $\hat G\hrho$ becomes the U(1) symmetry acting on the left replica space, and $\hrho \hat G^\dagger$ becomes the U(1) symmetry acting on the right space: 
\beqn 
&& \hat G \, \hrho \ \ \ra \ \ \left( \hat G_L |\Psi_0\rangle_L \right) \otimes |\Psi_0^\ast \rangle_R, \cr \cr 
&& \hrho \, \hat G^\dagger \ \ \ra \ \ |\Psi_0\rangle_L \otimes \left( \hat G_R^\ast |\Psi_0^\ast \rangle_R \right). 
\eeqn 
If we view and left and right spaces as spin-up and spin-down fermions, then the strong symmetry means that both spin-up and down fermions are conserved, and \beqn \hat G_L = \U(1)_\ua, \ \ \ \ \hat G_R = \U(1)_\da. \eeqn Or we can say that there is an effective ``charge-U(1)" and a ``spin-U(1)".

Under decoherence (density dephasing, say), or Lindbladian evolution, the left and right replica spaces become coupled (entangled): 
\beqn 
|\hrho_t \rrangle \sim e^{t \cL } |\hrho_0\rrangle, \eeqn where \beqn \cL &=& \sum_i \gamma \left( \hn_{i,L} \hn_{i,R} - \frac{1}{2}(\hn_{i,L}^2 + \hn_{i,R}^2) \right) \cr\cr 
&=& \sum_i - \frac{\gamma}{2} (\hn_{i,L} - \hn_{i,R})^2. 
\label{gamma}
\eeqn 
Therefore the Lindbladian $\cL$ acts as an ``attractive interaction" between the left and right spaces. And we know that for a spin-1/2 Fermi surface state, an attractive interaction can lead to long-range order of Cooper pairs $\hat{\Delta} = \hat{c}_{L} \hat{c}_{R}$:
\beqn
\lim_{x \ra \infty} \frac{\llangle \hrho | \hat{\Delta}_0 \ \hat{\Delta}_x^\dagger |\hrho \rrangle}{\llangle \hrho | \hrho \rrangle} \neq 0. 
\eeqn 
The long range Cooper pair correlation spontaneously breaks the $\U(1)_\ua \times \U(1)_\da$ (the strong U(1)) down to spin-U(1) (the weak U(1)). %therefore this Cooper pair condensate in the doubled space precisely corresponds to the SW-SSB.
If we write this Cooper pair correlation back in the density matrix form, it is exactly the R\'{e}nyi-2 correlator we discussed previously: 
\beqn 
C^{(2)}(x) &=&  \frac{\tr\left( \hrho \ \hat{c}_0 \hat{c}_x^\dagger \ \hrho \ \hat{c}_0^\dagger \hat{c}_x \right)}{\tr \hrho^2} = \frac{\llangle \hrho | \, \hat{\Delta}_0 \, \hat{\Delta}_x^\dagger \, |\hrho \rrangle}{\llangle \hrho | \hrho \rrangle} \cr\cr 
&& \lim_{x \ra \infty} C^{(2)}(x) \neq 0. 
\eeqn 
The doubled space representation of the density matrix provides an interpretation of SW-SSB using the standard notions in condensed matter: {\it The SW-SSB corresponds to the Cooper pair condensation and ODLRO in the doubled space.}

In fact, all the R\'{e}nyi correlators can be written as a correlation of a doubled state $|\hrho^{k/2}\rrangle$, by defining 
\beqn 
|\hrho^{k/2}\rrangle \sim \sum_{\bf{n}} p_{\bf{n}}^{k/2} \, | \bf{n} \rangle_L \otimes |\bf{n}\rangle_R. 
\eeqn

With the doubled space formula, the main results of our experiment can be naturally inferred: if we prepare a Fermi liquid state, dephasing becomes an inter-replica attractive interaction in the doubled space, which leads to inter-replica ``superconductivity", i.e. SW-SSB. By contrast, attractive interaction has little effect on a charge crystal, therefore a dephased crystalline insulator remains strongly symmetric, i.e. there is no inter-replica superconductivity, or SW-SSB. 

We summarize all possible scenarios of symmetry and symmetry breaking discussed in this overview in Table~\ref{tab:sixscenarios}.

\subsection{More Theoretical developments}

The notion of SW-SSB is multifaceted. We have so far discussed the symmetry aspect of SW-SSB, but it also has an intriguing information aspect connected to decodability and distinguishability, which is a unique aspect beyond the conventional symmetry breaking in equilibrium. In this section we briefly review the information aspect of SW-SSB through several examples.

The connection between SW-SSB and decodability was first discussed in the example of toric code, a paradigmatic topological order where information is stored nonlocally as loop operators. Under decoherence, the toric code topological order is corrupted with errors, nevertheless the stored information can be ``decoded" and recovered. Decoding is like an ``educated guess", based on the syndrome-measurements. Though different protocols of decoding, i.e. decoders, can vary in in their performances, it has been understood for a while that there exists a critical decoherence strength beyond which even the most optimal decoder would fail, meaning recovering the original information becomes impossible in the thermodynamic limit. 

This decodability phase transition was first discussed completely as an information transition, but it has been shown that it is precisely dual to a SW-SSB transition~\cite{lee2023,fan2024,bao2023}. This connection is based on the well-known duality in equilibrium between the $(2+1)d$ $Z_2$ gauge theory and $(2+1)d$ quantum Ising model, where the Ising degree of freedom corresponds to the $m$ (or $e$) anyon of the toric code, and the topological/confined phase is dual to the symmetric/symmetry-breaking phase of the Ising model. In the nonequilibrium setting, the decodable/undecodable phase is dual to the strongly-symmetric/SW-SSB phase of the Ising model. In the doubled space representation, the decodable/undecodable phase also corresponds to the insulator/``superconductor" of anyons: in the undecodable phase the anyons from the left (ket) and right (bra) replica spaces form a ``Cooper pair" bound state and condense. 

Another example of decodability was discussed recently in Ref.~\cite{hauser2026}, for systems with U(1) symmetry but no topological order. It was shown that if one starts with a charge crystal and evolves with diffusive dynamics, the charges are scrambled and hence the information of positions is gradually lost, and the total number of charge in a subregion becomes fuzzy. But one can still attempt to infer/decode the total number of charges in a subregion A, based on the measured ``syndrome", i.e. the charges in a neighbouring subregion B. But apparently the required size of B for the purpose of decoding grows with time, as it gets more challenging to decode at later time. It was shown that this decoding length scale (required size of B) closely tracks the correlation length of the R\'{e}nyi correlator which diagnoses the SW-SSB, and at the SW-SSB phase transition the decoding length scale should just diverge, meaning the information of charge number in A is completely lost, analogous to the loss of information stored in toric code.  

The third example of the connection to information is the interpretation of SW-SSB as a transition of distinguishability mentioned in the main text. For a fully dephased system, the density matrix is a mixture of different classical configurations, with probability given by the quantum wavefunction. As was discussed in Ref.~\cite{hauser2026} and shown in Eq.~\eqref{BC}, the R\'{e}nyi-1 correlator $C^{(1)}(x)$ that serves as the diagnosis of the SW-SSB, is also the Bhattacharyya Coefficient (BC) between two distributions $p$ and $p'$: $p$ is generated from the quantum wavefunction, while $p'$ is obtained from $p$ by moving one particle by distance $x$. The $C^{(1)}(x)$, or equivalently BC$(x)$ quantitatively measures the distinguishability between the two distributions: if $C^{(1)}(x)$ is very small, it means the particles are closely tied with their respective positions, and moving one particle by $x$ is very distinguishable; while if $C^{(1)}(x)$ is close to $1$, it means that the particles are scrambled enough and no longer tagged with their positions, hence moving particle by distance $x$ is no longer distinguishable. 

This information aspect not only conceptually enriches the scope of SW-SSB, it also provides a concrete protocol of measuring the R\'{e}nyi-1 correlator: according to the Bayes formula, BC is related to the success/failure rate of an attempt to distinguish two distributions. One can develop a tool (e.g. a neural-network) to perform the task of classifying density configurations into $p$ and $p'$, and the success/failure rate of this task should provide information of the R\'{e}nyi-1 correlator. 

%TC:endignore
%TC:endignore
\end{document}